%% file: main.tex
\documentclass[11pt,a4paper]{article}

\RequirePackage{ifpdf} 
\usepackage{amsmath} 
\usepackage{mathtools}

\usepackage{jheppub}
\usepackage{pstricks}
\usepackage[final]{pdfpages} 
\usepackage{ifpdf} 
\usepackage{slashed}
\usepackage{hyperref}

\usepackage{color} 
\usepackage{graphics}

\usepackage{etoolbox} 
\usepackage{fixmath}

\usepackage{amsfonts}

\usepackage{epstopdf}

\usepackage[normalem]{ulem}
\usepackage{framed}

\newcommand{\ep}{\epsilon}

\newcommand{\DD}{{\cal D}}
\newcommand{\nn}{\nonumber\\&}

\newcommand{\zb}{\bar{z}}
\newcommand{\zbb}{\bar{\bar{z}}}
\newcommand{\zbf}{\hat{\bar{z}}}
\newcommand{\zt}{\tilde{z}}
\newcommand{\ztt}{\tilde{\tilde{z}}}
\newcommand{\ztf}{\hat{\tilde{z}}}

\newcommand{\xxm}{\bar{t}}
\newcommand{\xxp}{\tilde{t}}
\newcommand{\xxt}{{\check{t}}}
\newcommand{\xxh}{\hat{\tilde{t}}}

\newcommand{\xyb}{\bar{w}}
\newcommand{\xyt}{\tilde{w}}

\newcommand{\rb}{\bar{\rho}}
\newcommand{\rrt}{\hat{\rho}}

\newcommand{\itwo}{i_2}
\newcommand{\mione}{i_1}

\newcommand{\GI}{G_I}
\newcommand{\sw}{s_{\scriptscriptstyle W}}
\newcommand{\cw}{c_{\scriptscriptstyle W}}

\newcommand{\be}{\begin{equation}}
\newcommand{\ee}{\end{equation}}
\newcommand{\bea}{\begin{eqnarray}}
\newcommand{\eea}{\end{eqnarray}}
\newcommand{\smallw}{{\scriptscriptstyle W}}
 
\newcommand{\mw}{m_\smallw}

\newcommand{\smallz}{{\scriptscriptstyle Z}}
\newcommand{\mz}{m_\smallz} 
\newcommand{\mzsq}{m_\smallz^2} 
 
\newcommand{\oa}{${\cal O}(\alpha)$\,} 
\newcommand{\oas}{${\cal O}(\alpha_s)$\,} 
\newcommand{\oasas}{${\cal O}(\alpha_s^2)$\,} 
\newcommand{\oaa}{${\cal O}(\alpha^2)$\,} 
\newcommand{\oaas}{${\cal O}(\alpha\alpha_s)$\,}

\newcommand{\eps}{\varepsilon} 
\newcommand{\muf}{\mu_F}
\newcommand{\mur}{\mu_R}
\newcommand{\sinw}{\sin\theta_W}

\title{On-shell $Z$ boson production at hadron colliders through \oaas}

\author[a]{Roberto Bonciani,}
\author[b]{Federico Buccioni,}
\author[c, d, e]{Narayan Rana,}
\author[d, e]{Alessandro Vicini}

\affiliation[a]{Dipartimento di Fisica, Universit\`{a} di Roma ``La Sapienza'' and INFN Sezione di Roma, Piazzale Aldo Moro 5, I-00185 Roma, Italy}
\affiliation[b]{Rudolf Peierls Centre for Theoretical Physics, Clarendon Laboratory, Parks Road, Oxford OX1 3PU, UK}
\affiliation[c]{Department of Physics, Indian Institute of Technology Kanpur, 208016 Kanpur, India}
\affiliation[d]{Dipartimento di Fisica ``Aldo Pontremoli'', University of Milano, Via Celoria 16, 20133 Milano, Italy}
\affiliation[e]{INFN Sezione di Milano, Via Celoria 16, 20133 Milano, Italy}

\emailAdd{roberto.bonciani@roma1.infn.it}
\emailAdd{federico.buccioni@physics.ox.ac.uk}
\emailAdd{narayan.rana@unimi.it}
\emailAdd{alessandro.vicini@mi.infn.it}

\abstract{The analytical expressions of the mixed QCD-EW corrections 
  to on-shell $Z$ boson inclusive production cross section at hadron colliders are presented, 
  together with computational details. The results are given in terms of polylogarithmic 
  functions and elliptic integrals. The impact on the prediction of the $Z$ boson production 
  total cross section is discussed, comparing different proton parton density sets.}

\colorlet{shadecolor}{gray!14}

\begin{document}

\preprint{\quad TIF-UNIMI-2021-20}

\allowdisplaybreaks[4]
\unitlength1cm
\maketitle
\flushbottom


\input{intro}

\input{cross-section}

\input{integrals}

\input{result}

\input{pheno}

\input{conclusion}

\bibliography{main} 
\bibliographystyle{JHEP}

\end{document}

%% file: intro.tex
\section{Introduction}
\label{sec:intro}

The production in hadronic collisions of a pair of leptons, each with large transverse momentum, is known as Drell-Yan (DY) process and it plays a fundamental role for our understanding of Quantum Chromodynamics (QCD) as the theory of the strong interactions. The lepton pair acts as a probe of the initial-state proton structure. It allows for the measurement of the proton collinear parton density functions (PDFs) and for the study of the QCD dynamics from the analysis of the lepton-pair transverse momentum distribution. The kinematical distributions of the final-state leptons allow for precision tests of the electroweak (EW) Standard Model (SM), with the determination of the weak mixing angle \cite{Aaltonen:2018dxj,ATLAS:2018gqq}, and of the masses $m_{W,Z}$ \cite{Group:2012gb,Aaboud:2017svj} of the $W$ and $Z$ bosons.

The production of an on-shell $Z$ boson represents an approximation of the full neutral-current DY process,
in that special kinematical configuration dominated by the $Z$-boson resonance. The total cross section of this process is an important theoretical benchmark constraining the proton PDFs, so that its prediction is a cornerstone in the precision physics program at hadron colliders.
This process is described in lowest order (LO) by quark-antiquark annihilation into a $Z$ boson via EW interaction. The evaluation of the next-to-leading order (NLO) \cite{Altarelli:1979ub}, next-to-next-to-leading order (NNLO) \cite{Hamberg:1990np,Catani:2009sm,Melnikov:2006kv,Harlander:2002wh}, and next-to-next-to-next-to-leading order (N$^3$LO) \cite{Ahmed:2014cla,Li:2014bfa,Duhr:2020seh,Duhr:2020sdp,Duhr:2021vwj} QCD corrections to the production of an on-shell gauge boson, supplemented by the resummation of the logarithmically enhanced terms due to soft gluon emission \cite{Sterman:1986aj,Catani:1989ne,Catani:1990rp,Moch:2005ky,Ravindran:2006cg,Catani:2014uta,H.:2020ecd,Camarda:2021ict}, allows for the accurate estimate of the total cross section, the reduction of the impact of QCD uncertainty and the precise assessment of its actual size (cfr. also Ref.\cite{Alioli:2016fum}). 

The best available QCD prediction for the inclusive production of a virtual gauge  boson includes up to N$^3$LO QCD corrections \cite{Duhr:2020seh,Duhr:2020sdp,Duhr:2021vwj}, in the $\gamma^*$ or $W$ cases respectively. It shows a dependence on the QCD renormalisation ($\mur$) and factorisation ($\muf$) scale choices at the sub-percent level for virtualities $Q>70$ GeV and at the percent level for smaller $Q$ values, with a stronger sensitivity to the choice of the factorisation scale. 

In this high-precision QCD framework, the inclusion of EW effects becomes mandatory. The NLO-EW corrections to the DY process have been computed in \cite{Baur:2001ze,CarloniCalame:2007cd,Arbuzov:2007db,Dittmaier:2009cr,Buonocore:2019puv} and are comparable in size to the NNLO-QCD effects. The theoretical uncertainty associated with missing higher-order EW corrections is formally at the NNLO-EW level and it is significantly reduced compared to the leading order (LO) case. Using different input parameters as a mean to estimate the size of missing higher-order EW effects, one finds that the LO variation is at the ${\cal O}(3.5\%)$ level, whereas the NLO-EW one is reduced down to the ${\cal O}(0.5\%)$ level.
Despite of these significant progresses, it is possible to observe the presence of some residual sources of uncertainty in the results listed above. The higher-order QCD predictions are only LO accurate from the point of view of the EW interaction, and thus they suffer from the uncertainty associated with the different choices of input parameters. If we consider the NNLO-QCD prediction, supplemented with the NLO-EW one, we find a scheme uncertainty at the ${\cal O}(0.88\%)$ level. This value is significant for any precision test, it is comparable to the residual QCD uncertainty, and in any case is slightly larger than the corresponding estimate based only on the LO+ NLO-EW results.
A specular discussion applies to the NLO-EW corrections, which are only LO from the point of view of the strong interaction. They suffer from large uncertainties under variations of the factorisation scale. The canonical $\muf$ variation by a factor 2 about its central value yields a change of the LO cross section by $\pm 18\%$ and, in turn, a change of the NLO-EW correction at the ${\cal O}(0.5\%)$ level. 
In order to increase the control on the theoretical error, it is therefore mandatory to include in the analysis the full mixed QCD-EW corrections, since they stabilize both the dependence on the QCD scales of the higher-order EW corrections and the dependence on the EW input parameters of the higher-order QCD corrections. 

The mixed corrections to the DY process in the resonance region have been studied in the so-called pole approximation in Refs.\cite{Dittmaier:2014qza,Dittmaier:2015rxo}, where QCD and EW effects are factorised between production and decay of the vector boson.
As far as the production of an on-shell $Z$ boson is concerned, the exact QCD-QED corrections have been considered in Refs.\cite{deFlorian:2018wcj,Cieri:2018sfk,Delto:2019ewv}, while in \cite{Bonciani:2016wya,Bonciani:2019nuy,Bonciani:2020tvf} we have computed the mixed QCD-EW effects fully analytically for the inclusive production.
In Ref.\cite{Buccioni:2020cfi}, the fully differential QCD$\times$EW effects have been presented, using a combination of analytical and numerical techniques. As for the full neutral-current DY, the QCD$\times$QED corrections to the production of a pair of neutrinos have been discussed in Ref.\cite{Cieri:2020ikq}, and the complete NNLO QCD-EW corrections for a charged lepton pair in the final state have been presented in \cite{Bonciani:2021zzf}.
Considering pole-approximation only for the virtual contributions, the QCD-EW corrections to the charged current DY has been obtained in \cite{Buonocore:2021rxx}.

In this paper, we explicitly present the analytical expressions and all the computational details to obtain the \oaas mixed QCD-EW corrections to on-shell $Z$ boson inclusive production cross section at hadron colliders. The results have been fully computed in analytical form and expressed in terms of polylogarithmic functions and elliptic integrals, requiring the evaluation of new two-loop master integrals (MIs) that were not available in the literature. The inclusion of the \oaas corrections increases the accuracy of the prediction and it reduces the impact of the residual theoretical uncertainties. The evaluation of the hadron-level cross section requires the convolution of the partonic results with proton PDFs. For consistency, the latter must satisfy DGLAP equations with also a QCD+QED evolution kernel. We comment on the accuracy of our predictions, compared to a standard analysis that includes only QCD corrections.

%% file: cross-section.tex
\section{The  \texorpdfstring{$Z$}{zb} boson production cross section}
\label{sec:xsec}

In this section, we discuss the theoretical framework, providing the details of perturbative contributions from various partonic channels which constitute the mixed QCD-EW corrections to the inclusive production cross section of the $Z$ boson at hadron colliders. We also discuss here the necessary steps to perform ultraviolet (UV) renormalisation and mass factorisation.

\subsection{The hadron-level cross section}

We organize the hadron-level inclusive $Z$ production cross section as a double perturbative expansion in the QCD and EW couplings, $\alpha_s$ and $\alpha$, respectively:
\begin{equation}
\sigma_H \equiv 
  \sigma(h_1 h_2\to Z+X) =
\sum_{m=0}^\infty
\sum_{n=0}^\infty\,\,
\alpha_s^m\, \alpha^n\,\sigma^{(m,n)}\, .
\label{eq:hse}
\end{equation}
The first term of this expansion, $\sigma^{(0,0)}$ is called Born cross section. All the contributions of \oas with respect to the Born form the NLO-QCD corrections $\sigma^{(1,0)}$, those of \oa are the NLO-EW corrections $\sigma^{(0,1)}$. At the second orders, the \oasas, \oaa and \oaas contributions are called the NNLO-QCD ($\sigma^{(2,0)}$), NNLO QCD-EW ($\sigma^{(1,1)}$) and NNLO EW ($\sigma^{(0,2)}$) corrections, respectively. In this paper we describe in detail the complete computation of $\sigma^{(1,1)}$.

The hadron-level cross section for the inclusive production of an on-shell $Z$ boson can be written, according to the factorisation theorem, as
\begin{equation}
\sigma_H =
\sum_{i,j=q,\bar q,g,\gamma}
\int_0^1 dx_1\,\int_0^1 dx_2\,
\tilde f_i^{h_1}(x_1)
\tilde f_j^{h_2}(x_2)\,
\tilde\sigma _{ij} \, .
\label{eq:upm}
\end{equation}
The bare PDFs, $\tilde f_i^{h}(x)$, describe the partonic content of the hadron $h$, where $i$ can be a quark ($q$), anti-quark ($\bar q$), gluon ($g$) or photon ($\gamma$). The partonic cross section $\tilde\sigma_{ij} \equiv \tilde\sigma(ij\to Z+X)$, describes the inclusive production of a $Z$ boson in the scattering of partons $i$ and $j$.

In the evaluation of the radiative corrections, the presence of UV and IR divergences is handled in dimensional regularisation, with $d=4-2\eps$ being the number of space-time dimensions. The individual matrix elements contain in general poles in $\eps$. The cancellation of the UV singularities takes place via UV renormalisation. The combination of virtual and soft real emission corrections to the same underlying process leads to a cancellation of the IR soft poles. According to the Kinoshita-Lee-Nauenberg (KLN) theorem \cite{Kinoshita:1962ur,Lee:1964is}, the combination of the different partonic cross sections contains only singularities due to the emission of collinear partons from the initial state. The latter are universal, they can be factorised and reabsorbed in the definition of the physical proton PDFs, $f_i(x,\muf)=\Gamma_{ik}(\muf) \tilde f_k(x)$, by means of the {mass factorisation} kernel $\Gamma$, defined at the factorisation scale $\muf$. As a consequence, Eq.~(\ref{eq:upm}) can be written as follows with the convolution represented by the symbol $\otimes$,
\begin{align}
\sigma_H &= \sum_{i,j=q,\bar q,g,\gamma}  \int_0^1 dx_1 \int_0^1 dx_2 \,
           f_i^{h_1}(x_1,\muf) f_j^{h_2}(x_2,\muf)\,
           \Gamma^{-1}_{ik}(\muf)\Gamma^{-1}_{jl}(\muf)\,  { \tilde\sigma_{kl} (x_1,x_2)} \, ,
\nonumber\\
&\equiv ~\sigma_0 \!\! \sum_{i,j=q,\bar q,g,\gamma}
f_i^{h_1}(\muf) \otimes f_j^{h_2}(\muf)\otimes\, \Delta_{ij}(\muf) \,.
\label{eq:physPDF}
\end{align}
The constant $\sigma_0$ is defined through the Born cross section $\sigma^{(0,0)}$,
\begin{equation}
\sigma^{(0,0)}  \equiv \sigma_0 c_q^{(0)} \delta(1-z) = \frac{4\sqrt{2} \pi}{N_C} G_\mu c_q^{(0)} \delta(1-z) \, .
\end{equation}
In the latter, $z\equiv\mzsq/\hat s$, $\sqrt{\hat s}$ is the partonic center of mass energy, $N_C$ is the number of colors, and $c_q^{(0)}=( C_{v,q}^2 + C_{a,q}^2 )$ is the combination of charges of the coupling of the $Z$ boson to a quark $q$. $C_{v,q}$ and $C_{a,q}$ are given by
\begin{equation}
 C_{v,q}=  \bigg( \frac{I_W^{(q)}}{2} -  \sinw Q_q \bigg) , \quad
 C_{a,q}=  \bigg( \frac{I_W^{(q)}}{2} \bigg) \, ,
 \label{vectandaxvect}
\end{equation}
with $I_W^{(q)}$ and $Q_q$ the third component of the weak isospin and the electric charge in units of the positron charge, respectively. $\theta_W$ is the weak mixing angle.

$\Delta_{ij}$ is the UV- and IR-finite partonic cross section for the partonic channel $ij$, expressed in units $\sigma_0$. In perturbation theory, $\Delta_{ij}$ is expanded in series of $\alpha_s$ and $\alpha$ as
\begin{equation}
 \Delta_{ij} = \sum_{m=0}^\infty \sum_{n=0}^\infty\,\, \alpha_s^m\, \alpha^n\, \Delta_{ij}^{(m,n)}\, . 
\end{equation}
%
The main results of this paper are the expressions of the corrections
$\Delta^{(1,1)}_{ij}$ with $i,j=q,\bar q,g,\gamma$.
In the study of the NNLO QCD-EW corrections, we encounter different gauge invariant subsets, characterised by the exchange of additional  photons or massive weak bosons $W$s or $Z$s. We reorganize accordingly the total \oaas correction, introducing
\begin{equation}
\alpha \alpha_s \sigma^{(1,1)} =  \sigma_0  \frac{\alpha}{4\pi} \frac{\alpha_s}{4 \pi} \left(
   \Delta^{(1,1)}_\gamma 
 + \frac{1}{\sw^2 \cw^2} \Delta^{(1,1)}_Z  
 + \frac{1}{\sw^2} \Delta^{(1,1)}_W
 \right) \, .
\label{eq:deltaqedew}
\end{equation}
We present in Section \ref{sec:results} the expressions of $\Delta^{(1,1)}_{\gamma}$, $\Delta^{(1,1)}_{Z}$, and $\Delta^{(1,1)}_{W}$.

\subsection{The partonic subprocesses}

\begin{figure}
\centering
\includegraphics[trim=0cm 21cm 0cm 0cm, clip=true, totalheight=4.5cm, angle=0]{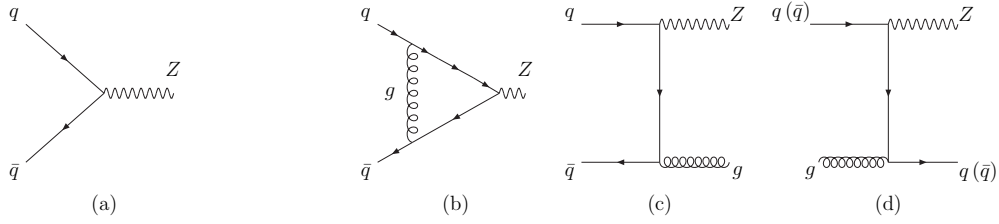}
\caption{\label{fig:treenloqcd} Feynman Diagrams contributing to the Born (a) and to the $\mathcal{O}(\alpha_S)$ corrections (b)--(d) to the production of a $Z$ boson in hadronic collisions. Crossed diagrams are not shown.}
\end{figure}

\begin{figure}
\centering
\includegraphics[trim=0cm 21cm 0cm 0cm, clip=true, totalheight=4.5cm, angle=0]{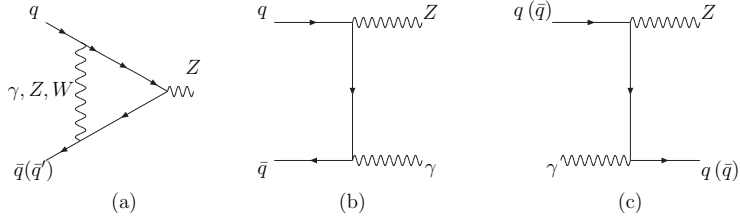}
\caption{\label{fig:nloew} Feynman Diagrams contributing to the $\mathcal{O}(\alpha)$ corrections to the production of a $Z$ boson in hadronic collisions. Crossed diagrams are not shown.}
\end{figure}

\begin{figure}
\includegraphics[trim=0cm 16cm 0cm 0cm, clip=true, totalheight=8.5cm, angle=0]{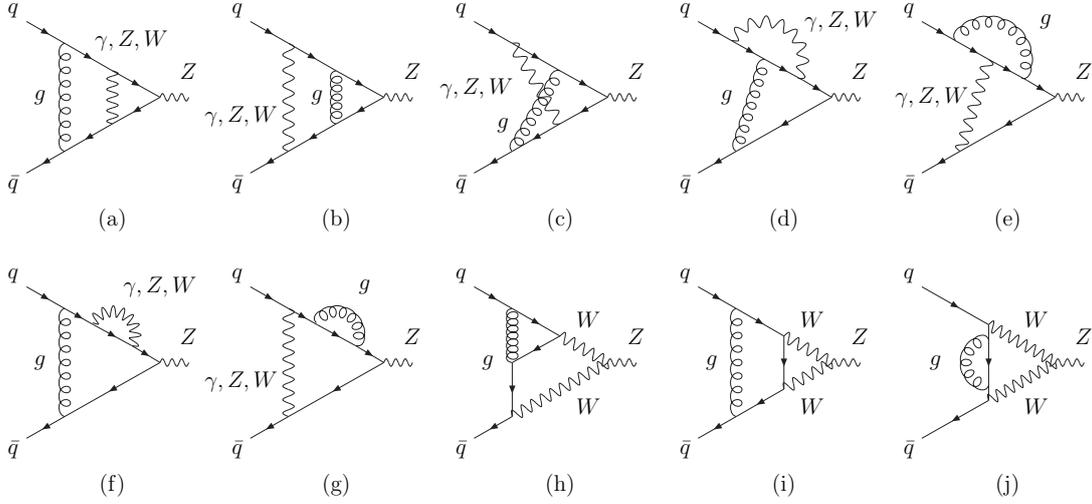}
\caption{\label{fig:twoloop} Two-loop Feynman Diagrams contributing to the $\mathcal{O}(\alpha \alpha_S)$ corrections to the production of a $Z$ boson in hadronic collisions, Eq.~(\ref{eq:2lvv}). Symmetric diagrams are not shown.}
\end{figure}

\begin{figure}
\centering
\includegraphics[trim=0cm 6cm 0cm 0cm, clip=true, totalheight=14.5cm, angle=0]{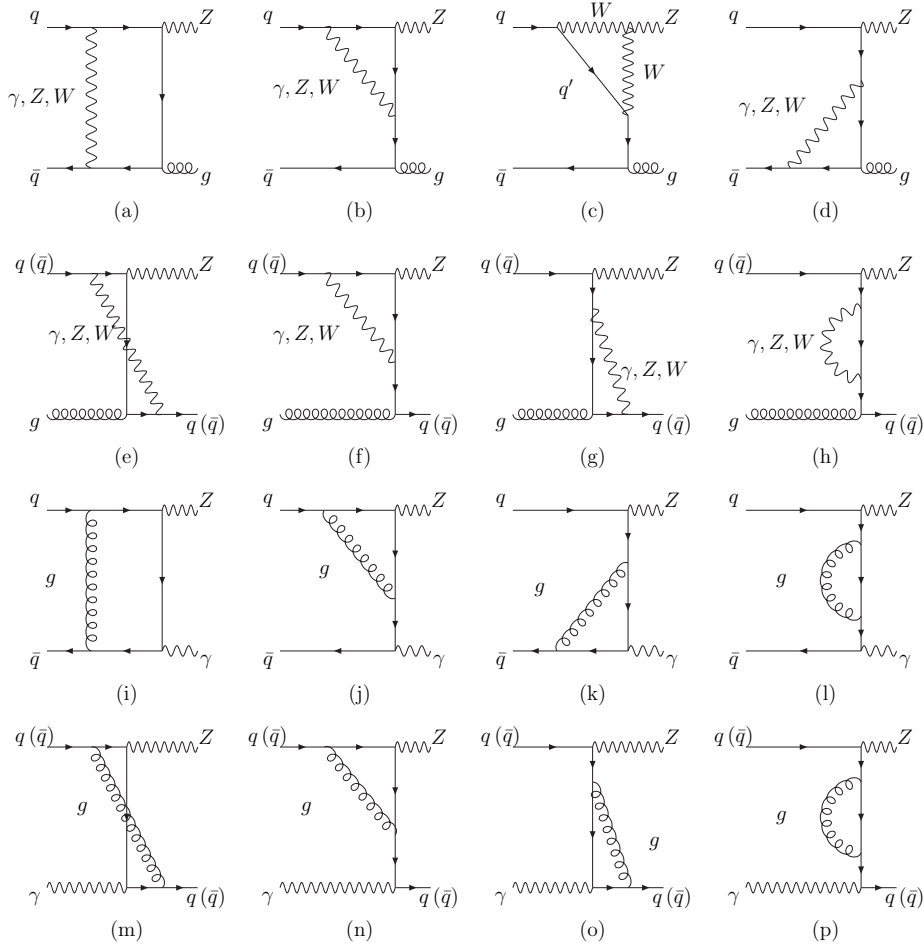}
\caption{\label{fig:realvirt} Examples of Feynman Diagrams contributing to the real-virtual corrections to the production of a $Z$ boson in hadronic collisions, Eqs.~(\ref{eq:2lrv}).}
\end{figure}

\begin{figure}
\centering
\includegraphics[trim=0cm 6cm 0cm 0cm, clip=true, totalheight=14.5cm, angle=0]{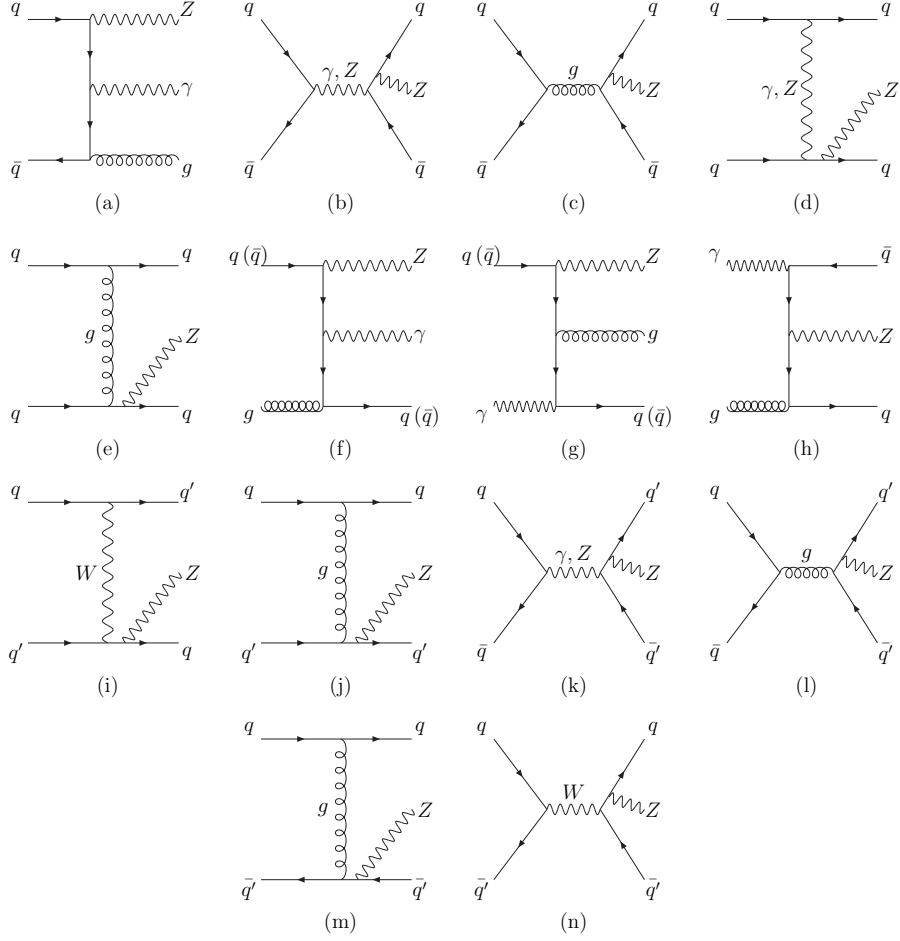}
\caption{\label{fig:doublereal} Examples of Feynman Diagrams contributing to the double real corrections to the production of a $Z$ boson in hadronic collisions, Eqs.~(\ref{eq:2ldr}).}
\end{figure}

The inclusive production cross section receives, order by order in perturbation theory, virtual as well as real emission corrections. The additional final-state partons present in the latter are completely integrated over their respective phase space.
The first term of the expansion $\sigma^{(0,0)}$ in Eq.~(\ref{eq:hse}), receives contributions from the single partonic process (see Fig.~\ref{fig:treenloqcd}~(a))
\begin{equation}
 q + \bar{q} \rightarrow Z \, .
\end{equation}
$\sigma^{(1,0)}$ in Eq.~(\ref{eq:hse}) receives contributions from the partonic processes
\begin{equation}
 q + \bar{q} \rightarrow Z \, , \quad
 q + \bar{q} \rightarrow Z + g \, , \quad 
 q + g \rightarrow Z + q \, .
\end{equation}
The first process receives NLO-QCD virtual corrections (Fig.~\ref{fig:treenloqcd}~(b)), while the other two
are evaluated at tree level, with the phase-space of the emitted parton ($g/q$) integrated out (Fig.~\ref{fig:treenloqcd}~(c),(d)). In case of $\sigma^{(0,1)}$, the processes are (see Fig.~\ref{fig:nloew})
\begin{equation}
 q + \bar{q} \rightarrow Z \,, \quad
 q + \bar{q} \rightarrow Z + \gamma \,, \quad 
 q + \gamma \rightarrow Z + q\,,
\end{equation}
where the first process receives NLO-EW virtual corrections (Fig.~\ref{fig:nloew}~(a)). 
At \oaas we have double-virtual, real-virtual and double-real contributions. Double-virtual corrections are two-loop contributions, with one gluon and one/two EW gauge bosons in the loop, to the partonic process (see Fig.~\ref{fig:twoloop})
\begin{equation}
 q + \bar{q} \rightarrow Z \, . 
\label{eq:2lvv}
\end{equation}
In the real-virtual contributions we find one virtual loop and one real-emitted particle. The processes which contribute to this group are (see Fig.~\ref{fig:realvirt})
\begin{equation}
 q + \bar{q} \rightarrow Z + g \, , \quad 
 q + g \rightarrow Z + q\,, \quad
 q + \bar{q} \rightarrow Z + \gamma \, , \quad 
 q + \gamma \rightarrow Z + q\, . 
\label{eq:2lrv}
\end{equation}
For the first two processes, the loop integral is of \oa (Fig.~\ref{fig:realvirt}~(a)--(h)), while for the others it is of \oas (Fig.~\ref{fig:realvirt}~(i)--(p)). 
At \oaas, the double-real contributions are with two real-emitted partons. Their amplitudes are evaluated at tree level. In the cases with two (anti)quarks in the initial and two (anti)quarks in the final state, in addition to the $Z$, the scattering is mediated by either a gluon or an EW boson, so that the respective contributions have a different proportionality in terms of $\alpha$ and $\alpha_s$. The interference between the two groups of contributions is of \oaas with respect to the Born process and is relevant for the calculation of $\sigma^{(1,1)}$. The complete list of processes is (see Fig.~\ref{fig:doublereal})
\bea
&&
 q + \bar{q} \rightarrow Z + g + \gamma \,, \quad
 q + \bar{q} \rightarrow Z + q + \bar{q} \,, \quad 
 q + q \rightarrow Z + q + q \,,  
 \nonumber\\
&&
 q + g \rightarrow Z + q + \gamma \,, \quad
 q + \gamma \rightarrow Z + g + q \,, \quad
 g + \gamma \rightarrow Z + q + \bar{q} \,,  
 \nonumber\\
&&
 q + q' \rightarrow Z + q + q' \,,  \quad
 q + \bar{q} \rightarrow Z + q' + \bar{q'} \,, \quad 
 q + \bar{q'} \rightarrow Z + q + \bar{q'} \,.
\label{eq:2ldr}
\eea
We denote with $q'$ a different quark flavour.

We analytically  compute the contribution at \oaas to the total cross section of each of the above listed processes, which can be presented as a Laurent expansion in $\eps$. {After the combination of all the degenerate states and mass factorisation, the singularities cancel and the remaining non-vanishing finite contributions are the factors $\Delta^{(1,1)}_{ij}$.} We present in Section \ref{sec:results} their analytical expressions, and provide them also in a {\tt Mathematica} ancillary file.

\subsubsection{Classification of the radiative corrections}
 
The \oa and \oaas corrections can be organised in a gauge invariant way into a photonic and two weak subsets.
The photonic subset ($\Delta_{\gamma}$) contains all contributions involving a photon i.e. a virtual photon in the loop or a real-emitted photon or a photon-initiated channel; at \oaas these are dubbed QCD-QED corrections.
The first weak subset ($\Delta_{Z}$) contains contributions from an additional $Z$ boson, either in the loop or as the intermediate propagator in the tree diagrams of double-real contributions. The other weak subset ($\Delta_{W}$) contains contributions from one or two $W$ bosons and includes Feynman diagrams with a non-abelian trilinear gauge boson vertex. This subdivision makes the computation, especially the mass factorisation, more organised and is useful for several intermediate checks.

\subsubsection{Ultraviolet renormalisation}
\label{sec:renormalisation}

The prediction of the hadron-level cross section requires to express the bare couplings and masses in terms of physical parameters via renormalisation. The choice of the background field gauge (BFG) \cite{Denner:1994xt} allows to restore the validity of $U(1)_{em}$-like Ward identities between the vertex corrections and the external quark wave function corrections in the full EW model. We explicitly verify these Ward identities at \oa and \oaas. We observe at \oa that the sum of the photonic correction in the vertex and external quark wave function corrections is UV finite. The same holds for the corrections with the exchange of one virtual $Z$ boson and, separately, for those with one or two $W$ bosons. An identical pattern takes place at \oaas. We are thus left with the discussion of the charge and external gauge boson wave function renormalisation.

If we choose to express $(g,g',v)$, the $SU(2)_L$ and $U(1)_Y$ couplings and the Higgs field vacuum expectation value, in terms of $(G_\mu,\mw,\mz)$ (dubbed $G_\mu$-scheme), where $G_\mu$ is the Fermi constant, then the weak charge renormalisation is achieved by the replacement
\be
\frac{g_0}{c_0} Z_{ZZ}^{1/2}
\to
\sqrt{4\sqrt{2} G_\mu \mz^2}\,
\left(1-\frac12 \Delta r +\frac12 \delta g_Z\right) \, .
\label{eq:chargeGmu}
\ee
We denote with a $0$ subscript all the bare quantities. We abbreviate with $\cw=\mw/\mz$ the cosinus of the weak mixing angle ($\sw^2=1-\cw^2$), and we define $ \delta g_Z\equiv \delta Z_{ZZ}+\delta e^2/e^2+(\sw^2-\cw^2)/(\cw^2) (\delta \sw^2/\sw^2)$ where $Z_{ZZ}=1+\delta Z_{ZZ}$ is the $ZZ$ wave function renormalisation constant. $\Delta r$ is a finite correction \cite{Sirlin:1980nh} expressing the relation between the Fermi constant and the muon decay amplitude, $\delta \sw^2=\cw^2\left(\delta\mz^2/\mz^2-\delta\mw^2/\mw^2\right)$, and $\delta e=e_0-e$, $\delta m_{W,Z}^2=m_{W,Z~0}^2-m_{W,Z}^2$ are the electric charge and gauge boson mass counterterms. The $\delta g_Z$ factor is, in the BFG, an UV finite correction and this fact yields a considerable simplification in the study of the impact of different input schemes. The $\Delta r$ parameter and the counterterms can be evaluated in perturbation theory and we keep terms of \oa and \oaas \cite{Kniehl:1989yc,Degrassi:2003rw}. For consistency, they have to be expressed in terms of $(G_\mu,\mw,\mz)$. In addition to the redefinition of the overall weak coupling, a second renormalisation correction modifies the vector coupling $v_q$ of the $Z$ boson to the quarks: $v_q = T_3^{(q)}-2 Q_q \sw^2 \to T_3^{(q)} - 2 Q_q (\sw^2+\delta \sw^2 +(\cw\,\sw/2)\, \delta Z_{AZ})$,
with $\delta Z_{AZ}$ the renormalisation constant of the $\gamma-Z$ mixing. In the BFG also this shift of $v_q$ is UV finite.

If we instead choose to relate $(g,g',v)$ to the $(\alpha,\mw,\mz)$ set of inputs (dubbed $\alpha(0)$-scheme),
the replacement of the overall coupling is $g_0/c_0\, Z_{ZZ}^{1/2}\to \sqrt{4\pi\alpha}/(\sw \cw) \left(1+\frac12 \delta g_Z\right)$, while the redefinition of the vector coupling remains the same as in the other scheme\footnote{
  An alternative input scheme, convenient to parameterise the $Z$ resonance,
  has been discussed in Ref.~\cite{Chiesa:2019nqb}.
}.

The $\alpha(0)$-scheme choice is historically \cite{Sirlin:1980nh,Denner:1991kt} one of the simplest EW renormalisation input schemes, but the low scale at which the fine structure constant is measured yields in turn the appearance of large logarithmic corrections in the perturbative expansion. The results depend on the value of the light-quark masses or, alternatively, on an experimental input $\Delta\alpha_{had}(\mz)=4\pi\left(\Pi_{\gamma\gamma}^{(5)}(\mzsq)-\Pi_{\gamma\gamma}^{(5)}(0) \right)$ \cite{Jegerlehner:2001wq,Jegerlehner:2018zrj,Keshavarzi:2018mgv,Davier:2019can} needed to evaluate the hadronic contribution to the running of the electromagnetic coupling at low scales, where $\Pi_{\gamma\gamma}^{(5)}(q^2)$ indicates the contribution of the first five light quark flavors to the photon vacuum polarisation at a scale $q^2$ (cfr. Ref.~\cite{Degrassi:2003rw}).

\subsubsection{Infrared singularities and mass factorisation}
 
Each UV-renormalised process, Eqs.~(\ref{eq:2lvv}-\ref{eq:2ldr}), is in general IR divergent, because of the exchange of a soft and/or collinear gluon or photon. Thanks to the KLN theorem, after combining all the partonic subprocesses and summing over all degenerate states, only initial-state collinear singularities are left. The latter are absorbed in the definition of the physical proton PDFs, by means of the mass factorisation kernel $\Gamma$. The subtraction kernels at \oaas are based on the splitting functions computed in Ref.\cite{deFlorian:2015ujt}. The photonic initial-state collinear singularities are reabsorbed in the definition of the physical proton PDFs and are resummed to all orders via the DGLAP evolution of the parton densities with a QED kernel. The consistent evaluation at \oaas of the hadron-level cross section requires a proton PDF set featuring DGLAP QED evolution, and, as already presented in Eq.~(\ref{eq:physPDF}), the inclusion of photon-induced subprocesses.

%% file: integrals.tex
\section{Computational details}
\label{sec:integrals}

In this section we present the details of the calculation. We first introduce our general strategy, which is based on the conversion of all the phase-space integrals into loop integrals, using the so-called reverse unitarity approach. We then discuss in detail the challenges posed by some new MIs with internal massive lines. We eventually provide a detailed account of how we deal with the appearing elliptic integrals.

\subsection{General strategy}
 \label{sec:generalstrategy}
The evaluation of the partonic cross-sections beyond LO requires the computation of Feynman loop integrals from virtual diagrams as well as two- and three-particle phase-space integrals arising from real emissions. In the very beginning of inclusive NNLO calculations, the phase-space integrals were performed using parametric and angular integration. However, to benefit from state-of-the-art techniques, developed for virtual integrals, we use the method of reverse unitarity~\cite{Anastasiou:2002yz,Anastasiou:2012kq} to convert the phase-space integrals into loop integrals. The reverse unitarity technique relies on the fact that the following replacement, Cutkosky rule,
\begin{equation}
 \delta(p^2-m^2) \rightarrow \frac{1}{2 \pi i} \bigg( \frac{1}{p^2-m^2+i 0_+} - \frac{1}{p^2-m^2-i 0_+} \bigg) 
\end{equation}
allows to convert the Dirac delta function in the phase-space measure of each final state particle into the difference of two propagators with opposite prescriptions for their imaginary part, where $0_+$ is an infinitesimal positive real number. The resulting integrals can then be studied with the help of standard techniques for virtual integrals.

The amplitudes that one needs to calculate can be classified as follows: two-loop $2\to 1$ virtual amplitudes and two-loop $2\to 2$ forward-scattering amplitudes with two- or three-particle cuts, stemming from the real-virtual and double-real corrections, respectively. There are up to two internal masses, which are in general complex valued. Once the Feynman diagrams contributing to these processes are generated using tools like {\sc Qgraf} \cite{Nogueira:1991ex} or {\sc FeynArts} \cite{Hahn:2000kx}, we compute the interference with the corresponding tree-level. We use in-house {\sc Form} \cite{Vermaseren:2000nd} or {\sc Mathematica} codes for this algebraic part of the computation. The interference term is expressed in terms of a large number of scalar integrals, that are not all independent, evaluated in $d$ space-time dimensions. Dimensionally regularised scalar integrals satisfy integration-by-parts (IBP) identities \cite{Tkachov:1981wb,Chetyrkin:1981qh,Laporta:2001dd}. These identities link different scalar integrals with each other and make possible the reduction of a large number of terms to a small set of independent quantities, called the MIs.
For the IBP reduction process, we have used the public programs {\sc Kira} \cite{Maierhofer:2017gsa}, {\sc LiteRed} \cite{Lee:2013mka, Lee:2012cn} and {\sc Reduze}2~\cite{vonManteuffel:2012np, Studerus:2009ye}.

The EW corrections entail the presence of masses in the loop propagators, specifically $m_Z$ and $m_W$. Accordingly, the integrals that have to be evaluated depend, in general, on three scales (or two dimensionless ratios). 
The fact that the two masses are numerically very close to each other can be used to reduce, effectively, the number of scales in the loop integrals, minimising the complexity of the calculation of the MIs. In fact, we can conveniently express the $W$ boson squared mass as
\begin{equation}
\label{eq:massexpansion}
m_W^2 = m_Z^2\left(1 - \xi \right), 
\quad\quad \text{with} \quad\quad
\xi = 1 - \frac{m_W^2}{m_Z^2} \simeq 0.2 \, .
\end{equation}
This makes it possible to expand the loop integrands, which contain $m_W^2$, as a Taylor series in the parameter $\xi$. Thus, loop integrals will only depend on $\hat{s}$ and $m_Z^2$, hence $z$, and the dependence on $m_W^2$ is formally confined into the coefficients of those integrals. 
The Taylor expansion in $\xi$ will generate loop propagators with higher powers, the so-called dotted propagators, without affecting the topology of the diagrams. Therefore, the set of MIs will not change.
The number of terms of such power expansion that have to be retained depends on the phenomenological accuracy we need for the evaluation of the corrections. 
%
After IBP reduction, the generic structure of the squared matrix element integrated over the inclusive phase space will then be
\begin{equation}
\int \mathrm{d}\Phi |{\cal M}|^2=\sum_{k=1}^{N} \sum_{j=0}^{n} \xi^j c_{k,j}(z,d)  I_k(z,d) \, ,
\end{equation}
where the $c_{k,j}$ are rational functions of the kinematical invariant $z$ and the number $d$ of space-time dimensions, while the $I_k(z,d)$ are the $N$ MIs of the process. In this work, we have considered only up to the $n=2$ order in the $\xi$ expansion\footnote{While for the evaluation of the MIs involved in the real-virtual and double real corrections such an expansion is indeed needed, the virtual corrections can be evaluated keeping the full dependence on $m_Z$ and $m_W$, as discussed in Section~\ref{VirtualCorr}.} and we will discuss the impact of this truncation in Section~\ref{se:phenomenology}.

The following step is to evaluate the MIs as a function of $z$ and $d$. We use the method of differential equations \cite{Kotikov:1990kg,Remiddi:1997ny,Gehrmann:1999as,Argeri:2007up,Henn:2013pwa,Henn:2014qga,Ablinger:2015tua,Ablinger:2018zwz} to solve the MIs. We differentiate one MI with respect to $z$ and we use the IBP identities on the differentiated output to express it as a combination of the MI itself and of other MIs. Applying this procedure on all the MIs of an integral family, we obtain a system of first-order linear differential equations, which can be solved given a set of boundary conditions (BCs).
In the current calculation the differential equations for most of the MIs, can be solved expressing the solution in terms of harmonic polylogarithms (HPLs), generalised harmonic polylogarithms\footnote{
Note that our definition of HPL with a positive letter is the GPL with a change of sign e.g. $H_1(z) = -G_1(z) \,, H_{i_1} (z) = - G_{i_1} (z)$.} (GPLs) \cite{Remiddi:1999ew,Goncharov:polylog,Goncharov:2001iea,Vollinga:2004sn} and cyclotomic HPLs \cite{Ablinger:2011te}. Three MIs in the evaluation of the double-real corrections, need elliptic extensions of such functions, known as elliptic polylogarithms.

\subsection{Evaluation of the full set of Master Integrals}
 
The MIs which appear in the reduction of the squared matrix element of the different partonic subprocesses can be classified in three groups associated to the two-loop virtual corrections, the two-particle phase-space integrals of the real-virtual corrections and the three-particle phase-space integrals of the double-real processes. All the two-loop virtual MIs were already available in the literature \cite{Fleischer:1997bw,Fleischer:1998nb,Aglietti:2003yc,Aglietti:2004tq,Aglietti:2004ki,Aglietti:2007as,Bonciani:2010ms,Kotikov:2007vr}. We recompute them using the method of differential equations, starting with a generic off-shell value of the $Z$ boson virtuality and then taking the on-shell limit of the results. The latter are expressed in terms of multiple zeta values and cyclotomic constants \cite{Ablinger:2011te}, which can be reduced to a set of independent constants as introduced in \cite{Henn:2015sem}.
The off-shell virtual integrals and most of their on-shell limits have been independently checked using {\sc {Fiesta}} \cite{Smirnov:2015mct}.

In the real-virtual and double-real corrections, the MIs with only massless internal lines were available from Ref.~\cite{Anastasiou:2012kq}. The new MIs with internal massive lines in real-virtual and double-real contributions are solved by employing again the differential equations technique. In these systems of differential equations, square root letters appear, making a direct solution cumbersome. To solve this problem, it is customary to perform a change of variables and to rationalize the square root factors. The presence, in some of the MIs under study, of multiple square roots makes it impossible to rationalize all the letters with one single variable transformation. In these cases, we exploit the linearity property of the integral operator and divide the MI into two (or more) subsystems. At this stage, we can choose for each subsystem a different transformation rule which rationalizes its specific letters. We trade the explicit dependence on $z$ of the full solution with the one on several new kinematical variables, but we obtain a simpler representation of the solution in terms of a compact alphabet as presented later in Eq.~(\ref{eq:alphabeth}).

Some of the MIs of the present study satisfy coupled differential equations and the problems of rationalising the letters and decoupling the equations are intertwined. We describe in the following example how the two issues can be simultaneously handled.

\subsubsection{Example : A subsystem of two real-virtual master integrals}

We consider two integrals $J_1 \equiv J_1(z)$ and $J_2 \equiv J_2(z)$ (see Fig.~\ref{mastersJ1J2}), which satisfy the following system of differential equations:
\begin{figure}
\centering
\includegraphics[trim=0cm 21cm 0cm 2cm, clip=true, totalheight=3.5cm, angle=0]{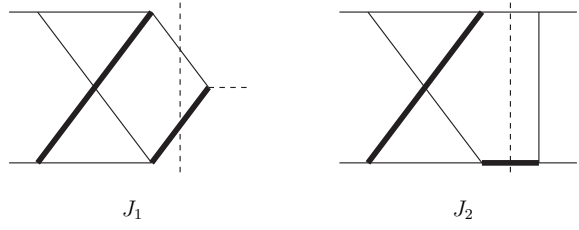}
\caption{\label{mastersJ1J2} Two MIs, $J_1$ and $J_2$, for the real-virtual corrections. Thin plain lines represent massless particles, while thick plain lines represent massive particles.}
\end{figure}
\begin{align}
 J_1' &= \frac{4 (d-3) z^2-d+6}{ z ( 1 + 4 z^2)} J_1 + r_1 (d, z) \,,
 \nonumber\\
 J_2' &= \frac{2}{z} J_2 
       + \frac{(2 d-9) (2 z+1)}{4 z^2+1} J_1 + r_2 (d, z)  \,.
\label{eq:deex1}       
\end{align}
The functions $r_{1,2}$ are the inhomogeneous part of the equations and contain rational functions and GPLs. With the transformation rule 
\begin{equation}
 z \rightarrow \frac{w}{1-w^2}
 \label{eq:exrule}
\end{equation}
and the relations
\begin{align}
 H_{0} (z) &= H_0 (w) + H_1 (w) - H_{-1} (w),
 \nonumber\\
 H_{1} (z) &= H_{i_1}(w) - H_{-i_2}(w) -  H_1 (w) + H_{-1} (w)
\end{align}
the first differential equation of Eq.~(\ref{eq:deex1}) can be rationalised and, with appropriate boundary conditions,
can be solved to obtain up to ${\cal O}(\eps^{-1})$ as
\begin{align*}
 J_1^{(-2)} =& 0 \,,
\nonumber\\
 J_1^{(-1)} =& \frac{w^2}{1-w^4} \Big(
i \pi  \big(
        3 H_0(w)
        +H_1(w)
        +H_{-1}(w)
\big)
-7 H_{0,0}(w)
-4 H_{0,1}(w)
+3 H_{0,-{i_2}}(w)
\nonumber\\
&
-3 H_{0,{i_1}}(w)
+4 H_{0,-1}(w)
-H_{1,0}(w)
-H_{1, {i_1}}(w)
-H_{-1,0}(w)
-H_{-1, {i_1}}(w)
+3 \zeta_2
\nonumber\\
&
+H_{1,-{i_2}}(w)
+H_{-1,-{i_2}}(w)
 \Big) \, .
\end{align*}
This allows us to readily obtain the solution for $J_2^{(-2)}$. However, the differential equation for $J_2^{(-1)}$ has
contributions to its non-homogeneous part from $r_2^{(-2)},r_2^{(-1)},J_2^{(-2)}$ as well as $J_1^{(-1)}$. The transformation rule in Eq.~(\ref{eq:exrule}) does not rationalize the former contributions, but it rather converts the linear kernels to much more involved polynomials. At this point the difficulty of solving the system for $J_{1,2}$ is evident, as cumbersome letters would appear in the GPLs, making the next integrations impossible. We reorganize the problem with the working hypothesis that, at two-loop level, all the coefficients of the poles in $\eps$ are expressible in terms of simple GPLs. In other words we expect to be able to find a combination of $J_1^{(n)}$ and $J_2^{(n)}$ such that their single pole can still be determined with elementary integrations.

With this goal, we define a new integral $J_0$
\begin{equation}
 J_0^{(n)} = \bigg( z-\frac{1}{2} \bigg) J_1^{(n)} +  J_2^{(n)}
 \label{eq:ex1combo}
\end{equation}
and we observe that $J_0^{(-1)}$ satisfies a linear differential equation independent of both $J_1^{(-1)}$ and $J_2^{(-1)}$. We solve it and obtain
\begin{align}
 J_0^{(-2)} =&
 z^2 \big( 
         H_0(z)
        +H_1(z)
        -i \pi
\big)\,,
 \nonumber\\
 J_0^{(-1)} =&
 -\frac{z^2}{2} \Big(
         2 H_{\frac{1}{2},0}(z)
        +2 H_{\frac{1}{2},1}(z)
        -7 H_{0,0}(z)
        -7 H_{0,1}(z)
        -10 H_{1,0}(z)
        -10 H_{1,1}(z)
\nonumber\\&
        +
                i \pi  \big(
                2 \ln (2)
                +3 H_0(z)
                +4 H_1(z)
        \big)
\Big) \, .
\end{align}
Consistently replacing each $J_2^{(n)}$ using Eq.~(\ref{eq:ex1combo}), both in the system of differential equations and in the matrix elements, we also remove the dependence of $J_1^{(0)}$. However, the contribution from $J_1^{(-1)}$, can not be eliminated, as expected. Hence, the issue of rationalisation remains in the full non-homogeneous part of $J_0^{(0)}$.
The non-homogeneous part of $J_0^{(0)}$ can be split in the sum of two terms, with or without a relation with the square root letter: we use the change of variables to $w$ to linearise only for the former, while we keep the variable $z$ for the latter. When we solve the non-homogeneous equation, we consider two separate integrals, with the integrand functions depending only on $w$ or only on $z$, with straightforward results in terms of GPLs, in the following form
\begin{align}
J_0^{(0)} =& z^2 \big(
-9 \zeta_3 + \cdots 
-2 H_{\frac{1}{2},\frac{1}{2},0}(z)
-2 H_{\frac{1}{2},\frac{1}{2},1}(z)
-5 H_{\frac{1}{2},0,0}(z)
-6 H_{\frac{1}{2},0,1}(z)
\nonumber\\&
+19 H_{1,1,1}(z)
+ \cdots 
\big)
+
\frac{w^2}{(1-w^2)^2} \big(
-3 H_{-1}(w) \zeta_2 + \cdots 
+H_{1,-1,0}(w)
+3 H_{1,0,{i_1}}(w)
\nonumber\\&
-3 H_{1,0,-{i_2}}(w)
+ \cdots  \big) \,.
\end{align}
The absence of the square root in the letters allows a smooth numerical evaluation. In the previous example, we find that a certain combination of MIs can be helpful to avoid the appearance of ``complicated'' GPLs in the intermediate steps of the solution of the system, under the assumption that they are not expected in the coefficients of the poles in $\eps$. This remark simplifies in turn the solution of the system for the finite part of the MIs. For example, the contributions from $J_1^{(0)}$ and $J_2^{(0)}$ do individually contain the problematic GPLs, which however would eventually cancel in the final result. Our direct evaluation of $J_0^{(0)}$ instead avoids these GPLs from the very beginning.

Apart from the square root letters, we also find three MIs in the double-real corrections for which the coupled sub-system of differential equations can not be factorised to first order, indicating that the MIs contain elliptic polylogarithms.
In the next sub-section we present the details of the computation of these MIs.

\subsubsection{Solving the master integrals with elliptic kernels}

\begin{figure}
\centering
\includegraphics[trim=0cm 21cm 0cm 2cm, clip=true, totalheight=3.5cm, angle=0]{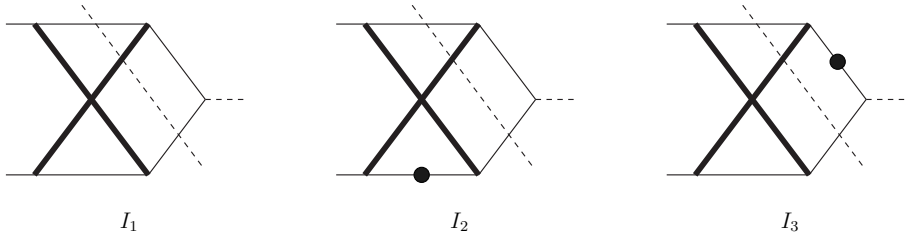}
\caption{\label{mastersELL} The three elliptic MIs, $I_1$, $I_2$ and $I_3$, contributing to the real-virtual corrections. Thin plain lines represent massless particles, while thick plain lines represent massive particles. A dot on a line represents the squared propagator.}
\end{figure}

The double real corrections receive contributions from scattering processes with quark and anti-quark both in the initial and final state, whose matrix elements include contributions with the exchange of either an EW boson or a gluon. The interference of these two terms is of \oaas\, and is relevant to our calculation. In these interferences, a non-trivial topology gives rise to three MIs $\{I_1,I_2,I_3\}$ (see Fig.~\ref{mastersELL}) which are of elliptic kind. The $3\times3$ system of differential equations is not first-order factorizable.
The homogeneous part of the system is given in each order of $\eps$ as
\begin{align*}
 \begin{split}
   d_z \left(
   \begin{array}{l}
      I_{1}^{(n)}\\
      I_{2}^{(n)}\\
      I_{3}^{(n)}
    \end{array}
    \right)&=
    \left[
      \begin{array}{ccc}
         \frac{4}{3 z}        & -\frac{2}{3 z}                            & \frac{2}{3} \frac{1}{z^2}            \\
          \frac{4}{3 (1+8 z)} & \frac{-3-31 z+16 z^2}{3 (-1+z) z (1+8 z)} & \frac{2 (5+4 z)}{3 (-1+z) z (1+8 z)} \\
         -\frac{2}{3 (1+8 z)} & \frac{5+4 z}{3 (-1+z) (1+8 z)}            & \frac{-11-46 z+48 z^2}{3 (-1+z) z (1+8 z)} \\
      \end{array}
      \right]
    \left(
    \begin{array}{l}
      I_{1}^{(n)}\\
      I_{2}^{(n)}\\
      I_{3}^{(n)}
    \end{array}
    \right)
  + \left(
    \begin{array}{l}
      R_{1}^{(n)}\\
      R_{2}^{(n)}\\
      R_{3}^{(n)}
    \end{array}
    \right) \, .
  \end{split}
\end{align*}
The homogeneous part of this system of equations is the same as the one studied for the corresponding virtual diagrams in Refs.~\cite{Aglietti:2007as,Broedel:2019hyg}. In those papers, the results are obtained in terms of elliptic integrals of the first kind and eMPLs, respectively. The current system can be solved order-by-order in $\eps$, and we expect to find only standard HPLs in the poles, while eMPLs will appear in the finite parts and higher orders in $\eps$.

\begin{figure}[h]
\centering
\includegraphics[height=3.5cm]{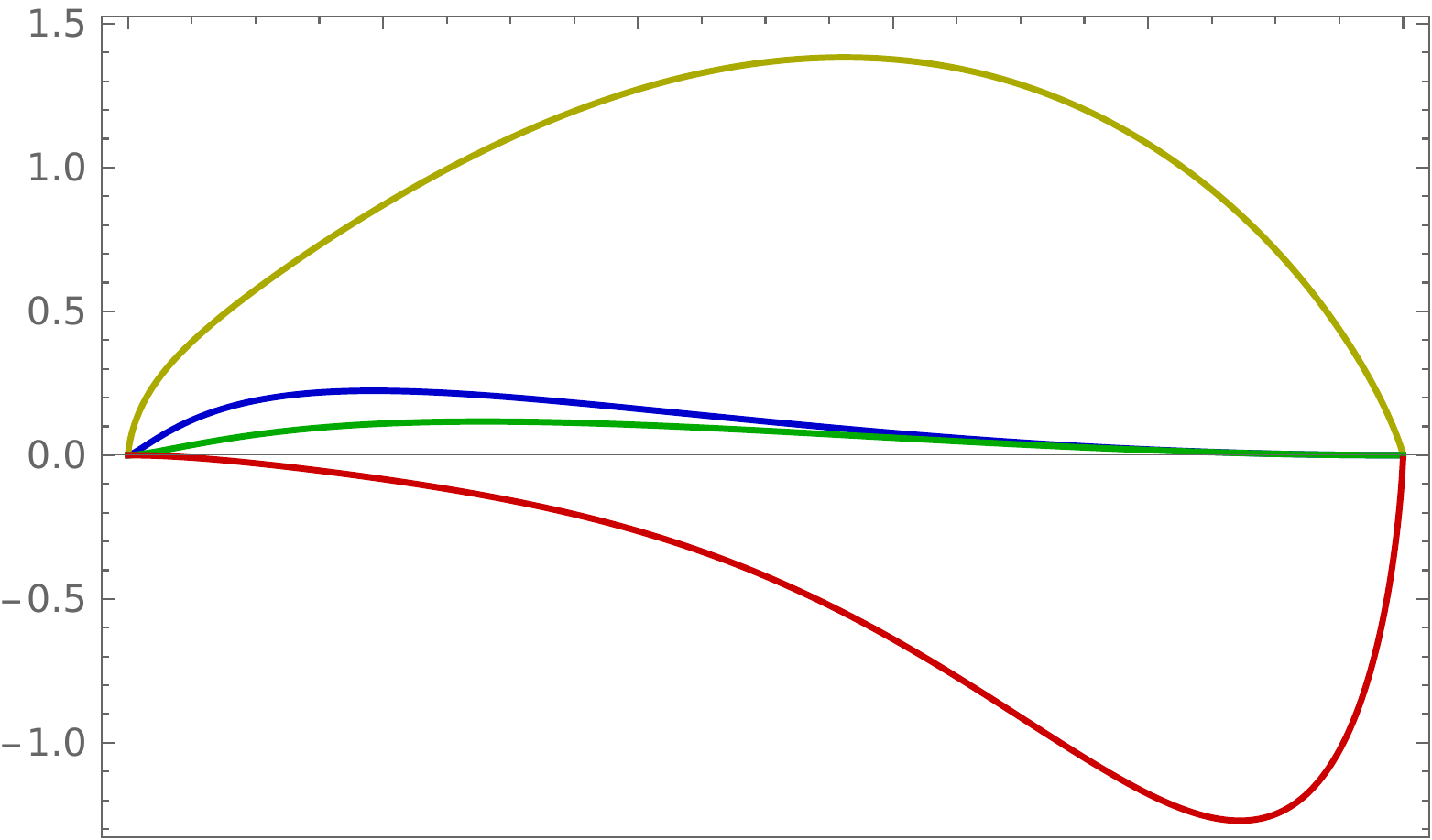}
\caption{The $\eps^0$ coefficient of the elliptic MIs i.e. $I_1^{(0)}$, $I_2^{(0)}$ and $I_3^{(0)}$ are presented 
in {\color{blue} blue}, {\color{yellow!80!black} yellow} and {\color{green!80!black} green}, respectively.
The $I_0^{(1)} 10^{-1}$ is shown in {\color{red} red}.
}
\label{fig:int}
\end{figure}

Formally, since the IBP reduction introduces a $1/\varepsilon$ factor in the coefficient of these integrals in the matrix elements, then we would need to evaluate $I_{1,2,3}$ up to ${\cal O}(\varepsilon)$, in order to compute all the finite corrections. For the same argument, we expect that individual contributions to the coefficient of the single pole of the matrix elements are expressed via eMPL. We formulate a physical Ansatz as in the previous Section, imposing the simpler polylogarithmic structure of the single pole of the matrix element, i.e. the absence of eMPLs in its final expression. We thus find the following combination of the elliptic MIs,
\begin{equation}
  I_0^{(n)} = z (1 + 2 z) I_1^{(n)} + z (1 - 4 z) I_2^{(n)} - (1 + 5 z) I_3^{(n)} \, .
\end{equation}
The differential equation of $I_0^{(-1)}$ and $I_0^{(0)}$ turns out to be linear and independent of $I_1^{(0)},I_2^{(0)}$ and $I_3^{(0)}$.
We solve it to obtain the following solution:
\begin{align}
 I_0^{(-1)} &= \frac{1}{2} z^2 (-1+4 z) H_0(z) \, ,
 \nonumber\\
 I_0^{(0)} &= \big(
        -\frac{5 z^2}{2}
        +\frac{6 z^4}{-1+z}
\big) H_{0,0}(z)
+2 z^2 (-1+4 z) H_{0,1}(z)
+2 (1-4 z) z^2 \zeta_2 \, ,
\end{align}
This new combination of MIs explicitly exhibits the absence of eMPL in their single pole and allows in turn a straightforward check of the pole cancellations in the cross section. Clearly, since the combination $I_0^{(1)}$ does not remain independent of $I_2^{(0)}, I_3^{(0)}$ we find eMPLs contributing to the finite part of the matrix elements.
For the numerical evaluation, a series representation of the integrals is equivalent to the formal solution via eMPLs. Hence, we solve $I_2^{(0)}$ and $I_3^{(0)}$ in Taylor series expansion around $z=0, 1/2, 1$. (See for instance \cite{Pozzorini:2005ff,Aglietti:2007as,Blumlein:2017dxp,Lee:2017qql,Lee:2018ojn,Bonciani:2018uvv,Blumlein:2019oas}).
$I_2^{(0)}$ and $I_3^{(0)}$ are regular in the whole range of $z$, while $I_0^{(1)}$ instead contains logarithmically enhanced terms. The latter yield a singular behaviour at a point which does not correspond to any physical threshold. These logarithms are thus expected to cancel against analogous terms stemming from other MIs. In order to achieve an exact analytical cancellation, we further elaborate the solution of $I_0^{(1)} \equiv I^{(0,1)}_{ell} = I_0^{(1.HPL)} + \delta^{(0,1)}_{ell}$, by splitting its expression in two parts: one with the closed form of the logarithmic dependence and one regular remainder given via a Taylor expansion.
We obtain 
\begin{align}
 I^{(0,1)}_{ell} =&
 -35 z^3 H_{-\frac{1}{2},0,0}(z)
-60 z^3 H_{-\frac{1}{2},0,1}(z)
+\frac{1}{2} z^2 \big(
        -3
        +86 z
        -24 z^2 \zb
\big) H_{0,0,0}(z)
-2 z^2 \big(
        5
\nonumber\\&        
        -35 z+42 z^2\big) \zb H_{0,0,1}(z)
-2 z^2 \big(
        2-12 z+7 z^2\big) \zb H_{0,1,0}(z)
-8 (1-4 z) z^2 H_{0,1,1}(z)
\nonumber\\&
-4 z^2 (1+z) (2+z) \zb H_{0,-1,0}(z)
-2 z^2 \big(
        1-3 z+8 z^2\big) \zb H_{1,0,0}(z)
+60 z^3 H_{-\frac{1}{2}}(z) \zeta_2
\nonumber\\&
+6 z^2 \big(
        1-12 z+14 z^2\big) \zb H_0(z) \zeta_2
+2 z^3 (-43+49 z) \zb \zeta_3
+  \delta^{(0,1)}_{ell} \,.
\end{align}
$\delta^{(0,1)}_{ell}$ has been obtained in Taylor series expansion. In the following,
we present $\delta^{(0,1)}_{ell}$ for the expansion around $y=1-z=0$ and $z=0$.
\begin{align}
\delta^{(0,1)}_{ell} =&
z^3 \bigg(
         35 H_{-\frac{1}{2},0,0}(1)
        +60 H_{-\frac{1}{2},0,1}(1)
        -60 H_{-\frac{1}{2}}(1) \zeta_2
\nonumber\\&                
        -20 y^2
        -\frac{1205 y^3}{54}
        -\frac{1535 y^4}{72}
        +\Big(
                10 y^2+\frac{130 y^3}{9}+\frac{295 y^4}{18} \Big) H_0(y) + {\cal O}(y^5)
\bigg) \, ,
\nonumber\\
= & \, 90 z^4
-360 z^5
+\frac{21745 z^6}{9}
-10 z^4 \big(
        -3-6 z+29 z^2\big) H_0(z)
-30 z^4 \big(
        7-29 z
\nonumber\\&        
        +198 z^2\big) H_{0,0}(z)
-45 z^3 \big(
        -1-12 z^2+72 z^3\big) H_{0,0,0}(z)
+\big(
        60 z^4
        -510 z^5
        +3860 z^6
\nonumber\\&        
        +360 z^5 (-1+6 z) H_0(z)
\big) \zeta_2
+20 z^3 \big(
        -1-27 z^2+162 z^3\big) \zeta_3
+ {\cal O} (z^7) \, .
\end{align}
$I_2^{(0)} \equiv I^{(2,0)}_{ell}$ and $I_3^{(0)} \equiv I^{(3,0)}_{ell}$ are obtained in the two series expansions as follows
\begin{align}
 I^{(2,0)}_{ell} = & \, 
3 y
-\frac{9 y^2}{4}
-\frac{13 y^3}{36}
-\frac{7 y^4}{72}
+\Big(
        -2 y+y^2+\frac{y^3}{3}+\frac{y^4}{6}
 \Big) H_0(y)
 + {\cal O}(y^5) \, , 
\nonumber\\
= & \, 
\frac{3}{4} z H_0(z)^2
-2 z^2 \big(
        -2+3 H_0(z)\big)
+z^3 \Big(
        -2 \big(
                4
                +12 \zeta_2
                +9 \zeta_3
        \big)
        -3 \big(
                -5
                +4 \zeta_2
        \big) H_0(z)
\nonumber\\&        
        +18 H_0(z)^2
        +3 H_0(z)^3
\Big)
+z^4 \Big(
        \frac{8}{9} \big(
                122
                +243 \zeta_2
                +162 \zeta_3
        \big)
        +2 \big(
                -55
                +48 \zeta_2
        \big) H_0(z)
\nonumber\\&        
        -162 H_0(z)^2
        -24 H_0(z)^3
\Big)
+z^5 \Big(
        \frac{1}{4} \big(
                -3671
                -7416 \zeta_2
                -5184 \zeta_3
        \big)
\nonumber\\&        
        -24 \big(
                -35
                +36 \zeta_2
        \big) H_0(z)
        +\frac{2781}{2} H_0(z)^2
        +216 H_0(z)^3
\Big)
+ {\cal O}(z^6) \, .
\end{align}
%
%
\begin{align}
 I^{(3,0)}_{ell} =& \, 
\frac{y^2}{2}-\frac{y^3}{4}-\frac{5 y^4}{48}
+ {\cal O}(y^5) \,,
\nonumber\\
= & \, 
z^2 \big(
        2 H_0(z)^2
        -2 \zeta_2
\big)
+z^3 \big(
        6
        +18 \zeta_2
        +18 \zeta_3
        +\big(
                -6
                +12 \zeta_2
        \big) H_0(z)
        -13 H_0(z)^2
\nonumber\\&        
        -3 H_0(z)^3
\big)
+z^4 \Big(
        -\frac{3}{2} \big(
                53
                +100 \zeta_2
                +72 \zeta_3
        \big)
        -3 \big(
                -17
                +24 \zeta_2
        \big) H_0(z)
        +113 H_0(z)^2
\nonumber\\&        
        +18 H_0(z)^3
\Big)
+z^5 \Big(
        \frac{1}{12} \big(
                7379
                +14256 \zeta_2
                +10368 \zeta_3
        \big)
        +\frac{1}{2} \big(
                -859
                +1152 \zeta_2
        \big) H_0(z)
\nonumber\\&        
        -\frac{1781}{2} H_0(z)^2
        -144 H_0(z)^3
\Big)
+ {\cal O}(z^6) \, .
\end{align} 
The series expanded $\delta^{(0,1)}_{ell}$, $I_2^{(0)}$ and $I_3^{(0)}$, are all regular and admit a Taylor expansion with fast convergence. This procedure provides a series representation of the elliptic functions present in the calculation. In Fig.~\ref{fig:int}, we present the numerical evaluation of the elliptic MIs for the whole range of $z$.

%% file: result.tex
\section{Analytical results}
\label{sec:results}

In this section we present our results. First, we introduce all the variables which appear in our calculation and several other necessary definitions. We then present the analytic finite partonic coefficients $\Delta^{(1,1)}_\gamma$, $\Delta^{(1,1)}_Z$ and $\Delta^{(1,1)}_W$.

\subsection{Preliminaries: variables, functions and abbreviations}

The partonic cross-section for the production of a $Z$ boson, integrated over the phase-space of the additional emitted partons, depends on the variables $s$, $m_Z$ and $m_W$ and on the top quark and Higgs boson masses $m_t$ and $m_H$ respectively only through UV renormalisation.
As described in Sec.~\ref{sec:generalstrategy}, we expand the virtual EW amplitudes containing $W$ bosons according to Eq.~(\ref{eq:massexpansion}).
For the purpose of clarity and compactness, we present here only results for the $n=0$ term in the $\xi$ expansion, i.e. $m_W^2 = m_Z^2$.

The Feynman integrals have a rich structure due to the presence of different internal thresholds. When the solution is given in terms of polylogarithmic functions\footnote{The solution is expressed in terms of Harmonic Polylogarithms (HPL), Generalised PolyLogarithms (GPL) or Cyclotomic HPLs.}, the internal structure of the integrals is displayed by the values of the weights and of the independent variable of the polylogarithms. The singular or branching points of the solutions are expressed in terms of a set of monomials, the letters, forming an alphabet. The presence in the letters of square-roots, also at multiple levels, can be avoided by appropriate transformations which rationalise the initial expressions. In the current calculation we adopt the following changes of variables:
\begin{equation} \label{eq:newvar}
 z = \frac{t}{(1+t)^2} = \frac{\rho}{(1 - \rho + \rho^2)} = \frac{w}{1-w^2} \, .
\end{equation}
After these manipulations, the resulting alphabet of the problem is given by:
\begin{equation}
 \Big\{ -1,-\frac{1}{2},0,\frac{1}{2},1,\{3,0\},\{3,1\},\{4,1\},\{6,0\},\{6,1\},i_1,-i_2\Big\} \,.
\label{eq:alphabeth}
\end{equation}
The set $\{-1,0,1\}$ is  the well-known alphabet for HPLs. $\{ \{3,0\},\{3,1\},\{4,1\},\{6,0\},\{6,1\} \}$
denotes the third-, fourth- and sixth-root of unity, which defines the cyclotomic HPLs. The set $\{ \{4,1\}, \mione,\itwo  \}$ appears in a group of diagrams expressed in terms of GPLs, where $i_1$ and $i_2$ are given by
\begin{equation}
 \mione = \frac{\sqrt{5}-1}{2} \equiv 0.618034 \ldots,
 \qquad
 \itwo = \frac{\sqrt{5}+1}{2} \equiv  1.618034 \ldots
\end{equation}
We enlist all our definitions below
\begin{align}
 &
 H_{-1} (z) = \int_{0}^z \frac{dx}{1+x} \,, \quad
 H_{0} (z) = \int_{0}^z \frac{dx}{x} \,, \quad
 H_{1} (z) = \int_{0}^z \frac{dx}{1-x} \,, 
 \nonumber\\&
 H_{-\frac{1}{2}} (z) = \int_{0}^z \frac{dx}{\frac{1}{2}+x} \,, \quad
 H_{\frac{1}{2}} (z) = \int_{0}^z \frac{dx}{\frac{1}{2}-x} \,, \quad
 \nonumber\\&
 H_{\{3,0\}} (z) = \int_{0}^z \frac{dx}{1+x+x^2} \,, \quad
 H_{\{3,1\}} (z) = \int_{0}^z \frac{x~dx}{1+x+x^2} \,, \quad
 \nonumber\\&
 H_{\{6,0\}} (z) = \int_{0}^z \frac{dx}{1-x+x^2} \,, \quad
 H_{\{6,1\}} (z) = \int_{0}^z \frac{x~dx}{1-x+x^2} \,, \quad
 \nonumber\\&
 H_{\{4,1\}} (z) = \int_{0}^z \frac{x~dx}{1+x^2} \,, \quad
 H_{i_1} (z) = \int_{0}^z \frac{dx}{i_1-x} \,, \quad
 H_{-i_2} (z) = \int_{0}^z \frac{dx}{i_2+x} \,.
\end{align}
In order to present the results in  compact form, we introduce the following abbreviations for all the polynomials which appear in the denominator of the expressions:
\bea
\zb  &\equiv&  \frac{1}{1-z},\,\,
\zbb  \equiv \frac{1}{1-2z},\,\,
\zbf  \equiv  \frac{1}{1-4z},\,\,
\zt   \equiv  \frac{1}{1+z},\,\,
\ztt  \equiv \frac{1}{1+2z},\,\,
\ztf  \equiv  \frac{1}{1+4z} \, , \\
\xxm &\equiv&  \frac{1}{1-t},\,\,
\xxp  \equiv  \frac{1}{1+t},\,\,
\xxh  \equiv  \frac{1}{1+t^2},\,\,
\xxt  \equiv  \frac{1}{1+t+t^2} \, , \\
\xyb   &\equiv&  \frac{1}{(1-w^2)^2},\,\,
\xyt  \equiv  \frac{1}{(1-w^4)} \, , \\
\rb  &\equiv&  \frac{1}{(1-\rho)},\,\,
\rrt   \equiv \frac{1}{(1-\rho+\rho^2)}\,\, .
\eea

\subsection{The virtual corrections \label{VirtualCorr}}

In this section we present the results for the sole virtual corrections.
The two-loop Feynman diagrams contributing to the renormalised $q \bar{q} Z$ vertex are shown in Fig.~\ref{fig:twoloop}.
We consider the interference with the tree-level and we express it in perturbative expansion of $\alpha$ and $\alpha_s$ as follows 
\begin{align}
 F = & \, \, 1 + \frac{\alpha_s}{4 \pi} F^{(1,0)}
        + \frac{\alpha}{4 \pi}  \left( F^{(0,1)}_\gamma  + \frac{1}{\sw^2 \cw^2} F^{(0,1)}_Z  + \frac{1}{\sw^2} F^{(0,1)}_W \right) 
\nonumber\\&  
        + \frac{\alpha}{4 \pi} \frac{\alpha_s}{4 \pi} \Big( F^{(1,1)}_\gamma  + \frac{1}{\sw^2 \cw^2} F^{(1,1)}_Z        
        + \frac{1}{\sw^2} F^{(1,1)}_W \Big) + .... \, .
\end{align}
The photonic contributions are denoted by $F_\gamma^{(m,n)}$ and they have been presented in \cite{deFlorian:2018wcj,Ajjath:2019ixh}.
Below, we present the renormalised contributions with one $Z$ boson or one/two $W$ boson/s in the loop, denoted by $F_Z^{(m,n)}$ and $F_W^{(m,n)}$, respectively.
We present here the results for the zeroth order in $\xi$.
We have also computed the completely general case with different masses
and compared it against the results available in the literature in~\cite{Kotikov:2007vr}, finding agreement.
We performed the calculation in the background field gauge. Therefore, we included the sole wave function renormalisation, which is sufficient to obtain UV finiteness.
%
In our expressions, the following constant $c_1$ \cite{Ablinger:2017hst} appears 
\begin{equation}
 c_1 = 12 \zeta_2 \ln^2 (2) + \ln^4 (2) + 24 \text{Li}_4 \Big( \frac{1}{2} \Big) \, .
\end{equation}
The one- and two-loop results for $F_Z^{(m,n)}$ and $F_W^{(m,n)}$, in the case of $u$-quark initial state, are in order:
\begin{align}
F_Z^{(0,1)} =& \, c_u^{(4)} \bigg[
 -\frac{11}{2}
+4 \zeta_2
- i \pi  \Big( 5-8 \ln (2) \Big)
\bigg] \,,
\\
F_Z^{(1,1)} =& \, C_F c_u^{(4)} \bigg[
\frac{1}{\ep^2} \Big\{
11
-8 \zeta_2
+ i \pi (10-16 \ln (2))
\Big\}
+ \frac{1}{\ep} \Big\{
52
-114 \zeta_2
-38 \zeta_3
+168 \zeta_2 \ln (2)
\nonumber\\&
+ i \pi \big(
        50
        -24 \zeta_2
        -28 \ln (2)
        +8 \ln ^2(2)
\big)
\Big\}
+ 195
-\frac{26 c_1 }{3}
-436 \zeta_2
+\frac{904}{5} \zeta_2^2
-\frac{229}{2} \zeta_3
\nonumber\\&
+366 \zeta_2 \ln (2)
-24 \zeta_2 \ln ^2(2)
+ {i \pi} \Big(
        261
        -118 \zeta_2
        -210 \zeta_3
        -38 \ln (2)
        +168 \zeta_2 \ln (2)
\nonumber\\&        
        +38 \ln ^2(2)
        -\frac{40}{3} \ln ^3(2)
\Big)
\bigg] \, ,
\\
 F_W^{(0,1)} =& \,
c_u^2 \Big( 
-\frac{7}{8}
+\frac{5}{4}  \frac{\pi}{\sqrt{3}}
-\zeta_2
\Big)
+ c_u c_d  \Big(
-\frac{1}{2}
-\frac{5}{4}   \frac{\pi}{\sqrt{3}}
+2 \zeta_2
- i \pi \Big( \frac{5}{4}-2 \ln (2) \Big)
\Big) \,,
\\
 F_W^{(1,1)} = & \,
C_F c_u^2 \bigg[
\frac{1}{\ep^2} \bigg\{
\frac{7}{4}
-\frac{5}{2}   \frac{\pi}{\sqrt{3}}
+2 \zeta_2
\bigg\}
+
\frac{1}{\ep} \bigg\{
\frac{19}{2}
-\frac{35}{4}   \frac{\pi}{\sqrt{3}}
+\frac{5}{2} \ln (3)  \frac{\pi}{\sqrt{3}}
+3 \zeta_2
-\frac{2}{3} \zeta_3
\nonumber\\&
-\frac{4}{3} \pi  G_I (0, r_2)
+10 \frac{1}{\sqrt{3}}  G_I (0, r_2)
+ i \pi \Big(
\frac{7}{4}
-\frac{5}{2}   \frac{\pi}{\sqrt{3}}
+2 \zeta_2
\Big)
\bigg\}
+ \bigg\{
\frac{143}{4}
-\frac{97}{4}   \frac{\pi}{\sqrt{3}}
\! + \!\frac{67}{6} \zeta_2
\nonumber\\&
+\frac{55}{9}   \frac{\pi}{\sqrt{3}} \zeta_2
-\frac{145}{9} \zeta_2^2
-\frac{1}{3} \zeta_3
-\frac{4}{3} \pi  G_I (0, {r_2})
+\frac{91}{2} \frac{1}{\sqrt{3}} {G_I} (0, {r_2})
+\frac{44}{3} {G_I} (0, {r_2})^2
\nonumber\\&
+2 \pi ~ {G_I} (0,1, {r_4})
+5 \sqrt{3} ~ {G_I} (0,1, {r_4})
+\frac{35}{4}   \frac{\pi}{\sqrt{3}} \ln (3)
+\frac{2}{3} \pi ~ {G_I} (0, {r_2}) \ln (3)
\nonumber\\&
-5 \frac{1}{\sqrt{3}} {G_I} (0, {r_2}) \ln (3)
-\frac{5}{4}   \frac{\pi}{\sqrt{3}} \ln ^2(3)
+ i \pi \bigg(
\frac{75}{2}
-\frac{35}{4}   \frac{\pi}{\sqrt{3}}
+3 \zeta_2
-\frac{74}{3} \zeta_3
\nonumber\\&
-\frac{4}{3} \pi  {G_I} (0, {r_2})
+10 \frac{1}{\sqrt{3}} {G_I} (0, {r_2})
+\frac{5}{2}   \frac{\pi}{\sqrt{3}} \ln (3)
\bigg)
\bigg\}
\bigg]
%
+ C_F c_u c_d \bigg[
\frac{1}{\ep^2} \bigg\{
1
+\frac{5}{2}   \frac{\pi}{\sqrt{3}}
-4 \zeta_2
\nonumber\\&
+ i \pi \bigg(
\frac{5}{2}-4 \log (2)
\bigg)
\bigg\}
+ \frac{1}{\ep} \bigg\{
\frac{7}{2}
+\frac{35}{4}   \frac{\pi}{\sqrt{3}}
-\frac{63}{2} \zeta_2
-\frac{53}{6} \zeta_3
+\frac{4}{3} \pi  {G_I} (0, {r_2})
\nonumber\\&
-10 \frac{1}{\sqrt{3}} {G_I} (0, {r_2})
+42 \zeta_2 \log (2)
-\frac{5}{2}   \frac{\pi}{\sqrt{3}} \log (3)
+ i \pi \bigg(
\frac{43}{4}
+\frac{5}{2}   \frac{\pi}{\sqrt{3}}
-8 \zeta_2
-7 \log (2)
\nonumber\\&
+2 \log ^2(2)
\bigg)
\bigg\}
+ \bigg\{
13
-\frac{13}{6} {c_1}
+\frac{97}{4}   \frac{\pi}{\sqrt{3}}
-\frac{721}{6} \zeta_2
-\frac{55}{9} \pi  \frac{\zeta_2}{\sqrt{3}}
+\frac{2759}{45} \zeta_2^2
-\frac{679}{24} \zeta_3
\nonumber\\&
+\frac{4}{3} \pi  {G_I} (0, {r_2})
-\frac{91}{2} \frac{1}{\sqrt{3}} {G_I} (0, {r_2})
-\frac{44}{3} {G_I} (0, {r_2})^2
-2 \pi  {G_I} (0,1, {r_4})
-5 \sqrt{3}  {G_I} (0,1, {r_4})
\nonumber\\&
+\frac{183}{2} \zeta_2 \ln (2)
-6 \zeta_2 \ln ^2(2)
-\frac{35}{4} \pi  \frac{1}{\sqrt{3}} \ln (3)
-\frac{2}{3} \pi  {G_I} (0, {r_2}) \ln (3)
\nonumber\\&
+ 5 \frac{1}{\sqrt{3}} {G_I} (0, {r_2}) \ln (3)
+ \frac{5}{4}   \frac{\pi}{\sqrt{3}} \ln ^2(3)
+ i \pi \bigg(
\frac{111}{4}
+\frac{35}{4}   \frac{\pi}{\sqrt{3}}
-\frac{65}{2} \zeta_2
-\frac{167}{6} \zeta_3
\nonumber\\&
+\frac{4}{3} \pi  {G_I} (0, {r_2})
-10 \frac{1}{\sqrt{3}} {G_I} (0, {r_2})
-\frac{19 \ln (2)}{2}
+42 \zeta_2 \ln (2)
+\frac{19}{2} \ln ^2(2)
-\frac{10}{3} \ln ^3(2)
\nonumber\\&
-\frac{5}{2}   \frac{\pi}{\sqrt{3}} \log (3)
\bigg)
\bigg\}
\bigg] \, ,
\end{align}
where we defined the following combinations of vector and axial-vector couplings:
\bea
c^{(4)}_{q} &=& C_{v,q}^4 + 6 C_{v,q}^2 C_{a,q}^2 + C_{a,q}^4 \, , 
\label{eq:cq4} \\
c_{q} &=& C_{v,q} + C_{a,q} \, ,
\label{eq:cq}
\eea
with $C_{v,q}$ and $C_{a,q}$ defined in Eq.~(\ref{vectandaxvect}) and $q \in (u,d)$.

\subsection{The partonic coefficients for QED}

The photonic part ($\Delta_{\gamma}^{(1,1)}$) of the total hadronic cross-section has been defined in Eq.~(\ref{eq:deltaqedew}) and receives contributions from several partonic channels which are convoluted with the physical proton PDFs as follows
\bea
 \Delta_{\gamma}^{(1,1)} &=& \sum_{q \in Q,\bar Q} f_q \otimes f_{\bar q} \otimes \Delta_{q\bar{q}}^{(1,1)}
+\sum_{q \in Q,\bar Q} f_q \otimes f_{q} \otimes \Delta_{qq}^{(1,1)} \nonumber\\
&& +\sum_{q \in Q,\bar Q} ( f_q \otimes f_{g} + f_g \otimes f_{q} ) \otimes \Delta_{qg}^{(1,1)}
+ \sum_{q \in Q,\bar Q} ( f_q \otimes f_{\gamma} + f_{\gamma} \otimes f_{q} ) \otimes \Delta_{q\gamma}^{(1,1)} \nonumber\\
&& + ( f_g \otimes f_{\gamma} + f_{\gamma} \otimes f_{g} ) \otimes \Delta_{g\gamma}^{(1,1)} \, .
\eea
The sums run over all flavours of quarks ($q$) and antiquarks ($\bar q$). The following combination of vector and axial-vector couplings, along with electric charge, appear in all these partonic channels with a photonic correction:
\be
c_q^{(2)} = Q_q^2 ~( C_{v,q}^2 + C_{a,q}^2 ) \, .
\label{couplingcombo}
\ee
In Eq.~(\ref{couplingcombo}) $q$ can be an up- or a down-type quark.
The partonic cross-section for all the channels are given below. We have set the factorisation scale $\mu_F=m_Z$ in presenting these results.
\begin{align}
\Delta_{q\bar{q}}^{(1,1)} =& \, c_q^{(2)} C_F \Big\{
\delta (1-z) \Big(
\frac{511}{2}
-140 \zeta_2
+\frac{16}{5} \zeta_2^2
-120 \zeta_3
\Big)
+ 
512 \zeta_3 \DD_0
- \DD_1 \Big(
512
+256 \zeta_2
\Big)
\nn
+ 256 \DD_3
+ \zeta_2 \Big(
192
-\frac{48}{1-z}
-176 z
-96 z^2
+\Big(
        16
        +\frac{64}{1-z}
        +176 z
        +96 z^2
\Big) H_0(z)
\nn
+\Big(
        -80
        -\frac{64}{1-z}
        -80 z
\Big) H_1(z)
-96 (1+z)^2 H_{-1}(z)
\Big)
+ 16 \zeta_3 \Big(
-3
-\frac{20}{1-z}
\nn
+13 z
+8 z^2
\Big)
+448
-328 z
-104 z^2
+\Big(
        64
        +\frac{320}{1-z}
        +152 z
\Big) H_0(z)
-8 (32
\nn
+31 z) H_1(z)
+\Big(
        96
        +\frac{120}{1-z}
        -240 z^2
\Big) H_{0,0}(z)
-\Big(
         64
        -\frac{96}{1-z}
        -288 z
\Big) H_{0,1}(z)
\nn
+\Big(
        -48
        +\frac{48}{1-z}
        +144 z
        -144 z^2
\Big) H_{1,0}(z)
+256 (-1+z) H_{1,1}(z)
\nn
+96 (1+z)^2 H_{-1,0}(z)
+\Big(
        184
        -\frac{160}{1-z}
        +280 z
        +32 z^2
\Big) H_{0,0,0}(z)
+\Big(
        320
        -\frac{448}{1-z}
\nn        
        +320 z
\Big) H_{0,0,1}(z)
+\Big(
        496
        -\frac{640}{1-z}
        +624 z
        +64 z^2
\Big) H_{0,1,0}(z)
+\Big(
        624
        -\frac{992}{1-z}
\nn
        +624 z
\Big) H_{0,1,1}(z)
+64 (1+z)^2 H_{0,-1,0}(z)
+\Big(
        304
        -\frac{416}{1-z}
        +304 z
\Big) H_{1,0,0}(z)
\nn
+\Big(
        512
        -\frac{960}{1-z}
        +512 z
\Big) H_{1,0,1}(z)
+\Big(
        560
        -\frac{1024}{1-z}
        +560 z
\Big) H_{1,1,0}(z)
\nn
+768 (1+z) H_{1,1,1}(z)
+160 (1+z)^2 H_{-1,0,0}(z)
-192 (1+z)^2 H_{-1,-1,0}(z)
\Big\} \, . \\
\Delta_{qg}^{(1,1)} =& \, c_q^{(2)} \Big\{
\zeta_2 \Big(
1
-20 z
+16 z^2
+2 \big(
        7-14 z+24 z^2\big) H_0(z)
+16 \big(
        1-2 z+2 z^2\big) H_1(z)
\Big)
\nn
+
\zeta_3 \Big( 26-52 z+68 z^2 \Big)
-\frac{1}{4} \Big( 157-442 z + 305 z^2 \Big)
-\frac{1}{2} \big(
        31-201 z+174 z^2\big) H_0(z)
\nn
+\big(
        -26+135 z-88 z^2\big) H_1(z)
+\frac{1}{2} \big(
        -11+60 z-4 z^2\big) H_{0,0}(z)
-2 \big(
        7-48 z
\nn        
        +48 z^2\big) H_{0,1}(z)
+\big(
        -23+72 z-56 z^2\big) H_{1,0}(z)
-2 \big(
        23-80 z+63 z^2\big) H_{1,1}(z)
\nn
-8 (1+z) (-1+3 z) H_{-1,0}(z)
+\big(
        -17+34 z-52 z^2\big) H_{0,0,0}(z)
-8 \big(
        3-6 z
\nn        
        +10 z^2\big) H_{0,0,1}(z)
-2 \big(
        11-22 z+40 z^2\big) H_{0,1,0}(z)
-6 \big(
        7-14 z+22 z^2\big) H_{0,1,1}(z)
\nn
+8 \big(
        1-2 z+2 z^2\big) H_{0,-1,0}(z)
-2 \big(
        -1+2 z+6 z^2\big) H_{1,0,0}(z)
-4 \big(
        9-18 z
\nn        
        +20 z^2\big) H_{1,0,1}(z)
-4 \big(
        7-14 z+16 z^2\big) H_{1,1,0}(z)
-70 \big(
        1-2 z+2 z^2\big) H_{1,1,1}(z)
\Big\}
%
\\
\Delta_{q\gamma}^{(1,1)} =& \,  2 C_A C_F \Delta_{qg}^{(1,1)} 
\\
\Delta_{qq}^{(1,1)} = & \, c_q^{(2)} C_F \Big\{
\zeta_2 \Big(
64
+144 z
-48 z^2
+\Big(
        -160
        +192 z
        -32 z^2
        +\frac{288}{1+z}
\Big) H_0(z)
\nn
-32 (1-z)^2 H_1(z)
+\Big(
        192
        -192 z
        -\frac{448}{1+z}
\Big) H_{-1}(z)
\Big)
- 16 \zeta_3 \Big(
 11
-11 z
-\frac{26}{1+z}
\Big)
\nn
-196
+248 z
-52 z^2
+8 (-16+13 z) H_0(z)
+128 (-1+z) H_1(z)
\nn
-16 \big(
        5+3 z^2\big) H_{0,0}(z)
-64 (1+z) H_{0,1}(z)
+16 (1-z) (-1+3 z) H_{1,0}(z)
\nn
+32 (1+z) H_{-1,0}(z)
+\Big(
        16
        +16 z
        -32 z^2
        -\frac{128}{1+z}
\Big) H_{0,0,0}(z)
\nn
+\Big(
        64
        -64 z
        -\frac{128}{1+z}
\Big) H_{0,0,1}(z)
-\Big(
         64
        -96 z
        +32 z^2
        -\frac{64}{1+z}
\Big) H_{0,1,0}(z)
\nn
-\Big(
        +64
        -64 z
        -\frac{192}{1+z}
\Big) H_{0,-1,0}(z)
-32 (1-z)^2 H_{1,0,0}(z)
-32 (1-z)^2 H_{1,1,0}(z)
\nn
-\Big(
         192
        -192 z
        -\frac{448}{1+z}
\Big) H_{-1,0,0}(z)
-\Big(
         128
        -128 z
        -\frac{256}{1+z}
\Big) H_{-1,0,1}(z)
\nn
+\Big(
        128
        -128 z
        -\frac{384}{1+z}
\Big) H_{-1,-1,0}(z)
\Big\} \,.
%
\\
\Delta_{g\gamma}^{(1,1)} = & \, C_A \Big(\sum_q c_q^{(2)} \Big) \Big\{
\zeta_2 \Big(
40 z
+16 z^2
+16 \big(
        1+3 z+3 z^2\big) H_0(z)
-16 (1+z)^2 H_{-1}(z)
\Big)
\nn
+ \zeta_3 \Big(
8-48 z+32 z^2
\Big)
-64
-132 z
+196 z^2
+2 \big(
        -23-64 z+105 z^2\big) H_0(z)
\nn        
+4 (-1+z) (7+67 z) H_1(z)
-8 (1+z) (3+4 z) H_{0,0}(z)
-8 \big(
        1+8 z-4 z^2\big) H_{0,1}(z)
\nn        
+48 (-1+z) (1+3 z) H_{1,0}(z)
+64 (-1+z) (1+3 z) H_{1,1}(z)
+16 (1+z) H_{-1,0}(z)
\nn
-8 \big(
        3+8 z+8 z^2\big) H_{0,0,0}(z)
-16 (1+2 z)^2 H_{0,0,1}(z)
-24 (1+2 z)^2 H_{0,1,0}(z)
\nn
-32 (1+2 z)^2 H_{0,1,1}(z)
+16 \big(
        1+2 z+2 z^2\big) H_{0,-1,0}(z)
-16 (-1+z)^2 H_{1,0,0}(z)
\nn
+48 (1+z)^2 H_{-1,0,0}(z)
-32 (1+z)^2 H_{-1,-1,0}(z)
\Big\} \, .
\end{align}
For completeness, we report the one-loop results.
\begin{align}
\Delta_{q\bar{q}}^{(0,1)} = & \, c_q^{(2)} \Big\{
\delta (1-z)  \big(
        -16
        +8 \zeta_2
\big)
+16 \DD_1
- \frac{4 (1+z^2)}{(1-z)} H_{0}(z)
+8 (1+z) H_{0}(z)
\Big\} \, .
\\
\Delta_{q\gamma}^{(0,1)} = & \, c_q^{(2)} C_A \Big\{
  (1-z) (1+7 z)
- 2 \big(1-2 z+2 z^2\big) \Big( H_{0}(z) + 2 H_{1}(z) \Big)
\Big\} \, .
\end{align}

\subsection{The partonic coefficients for Z}
 
$\Delta_{Z}^{(1,1)}$ represents the contribution to the total hadronic cross-section with a single internal $Z$ boson.
$\Delta_{Z}^{(1,1)}$ has the following dependency on the partonic cross-sections convoluted with PDFs
\begin{align}
 \Delta_{Z}^{(1,1)} = & \,
 \sum_{q \in Q,\bar Q} f_q \otimes f_{\bar q} \otimes \Delta_{q\bar{q},Z}^{(1,1)}
+\sum_{q \in Q,\bar Q} ( f_q \otimes f_{g} + f_g \otimes f_{q} ) \otimes \Delta_{qg,Z}^{(1,1)} 
\nn
 + \sum_{q \in Q,\bar Q} f_q \otimes f_{q} \otimes \Delta_{qq,Z}^{(1,1)} \,.
\end{align}
%

Below we present the partonic cross-sections. To renormalize UV divergences, only the quark wave function renormalisation has been performed in these results. The $q\bar{q}$ initiated partonic cross-section is given by
\begin{align}
\Delta_{q\bar{q},Z}^{(1,1)} =& \,  c^{(4)}_q C_F \Big\{
\delta (1-z) \Big(
204
-256 {\rm Li}_4 \Big( \frac{1}{2} \Big)
-\frac{32 \ln^4 (2)}{3}
-172 \zeta_2
+144 \ln (2) \zeta_2
\nn
 -128 \ln^2 (2) \zeta_2
+\frac{384}{5} \zeta_2^2
-56 \zeta_3
\Big)
+ \DD_1 \Big(
-176
+128 \zeta_2
\Big)
-16
+4 z
-8 \zb
+\big(
        -28
\nn 
         -68 z
        +80 \zb
        -8 \zb^2
\big) H_0(z)
-\big(
         88
        +88 z
        -(32+32 z) \zeta_2
\big) H_1(z)
+8 \big(
        5+15 z+4 z^2
\nn 
         +2 z^3\big) H_{0,0}(z)
+8 \big(
        -2+8 z+3 z^2+z^3\big) H_{1,0}(z)
-8 (1+z) \big(
        7+z^2\big) H_{-1,0}(z)
\nn
 -(224
+160 z
-256 \zb
) H_{0,0,0}(z)
-(112
+80 z
-128 \zb
) H_{0,1,0}(z)
+(112
+80 z
\nn
 -128 \zb
) H_{0,-1,0}(z)
-16 (5-z) H_{1,0,0}(z)
-16 (3+z) H_{1,1,0}(z)
+32 (1-z) H_{1,-1,0}(z)
\nn
 +(192
+64 z
-256 \zb
) H_{-1,0,0}(z)
+(96
+32 z
-128 \zb
) H_{-1,1,0}(z)
+(-96
-32 z
\nn
 +128 \zb
) H_{-1,-1,0}(z)
-\big(
         64
        -32 z
        -40 z^2
        -8 z^3
        + \ln (2) (96
        +96 z
        -192 \zb
        )
\big) \zeta_2
\nn
 +(48
+16 z
-64 \zb
) H_{-1}(z) \zeta_2
+16 (-1+7 \zb) \zeta_3
%
%
+ \frac{\delta_{q\bar{q},Z}}{z}
\Big\}
\, ,
\end{align}
where we used the combination of vector and axial-vector couplings $c^{(4)}_{q}$ defined in Eq.~(\ref{eq:cq4}), and where $\delta_{q\bar{q},Z}$ stems from the double-real channel $q\bar{q} \rightarrow q \bar{q} Z$, which does not produce any singularities. It is given as 
\begin{align}
\delta_{q\bar{q},Z} =&  \,
\frac{8 (1+2 z)^2}{3 z^2} I^{(0,1)}_{ell}
+\frac{8 {\ztt}}{3 z^2} \big(
        7+77 z+264 z^2+362 z^3+136 z^4\big) I^{(3,0)}_{ell}
+\frac{8 {\ztt}}{3 z} \big(
        -7-32 z
\nn 
         +6 z^2+232 z^3+224 z^4\big) I^{(2,0)}_{ell}
+ 2 z {\zb} {\zbf} \big(
        281-1537 z+1829 z^2-553 z^3-36 z^4-32 z^5\big)
\nn
 -4 z {\zb}^2 {\zbf}^2 \big(
        -91+803 z-2037 z^2+890 z^3+987 z^4-612 z^5-112 z^6+64 z^7\big) H_0(z)
\nn
 -\frac{4}{3} {\zb}^2 {\zbb} {\zbf}^2 {\ztt} \big(
        35-459 z+1687 z^2+434 z^3-10524 z^4-248 z^5+41084 z^6-14160 z^7
\nn
 -59984 z^8+40320 z^9+3840 z^{10}\big) H_{0,0}(z)
+\frac{16 {\ztt}}{3} \big(
        -7-32 z+6 z^2+232 z^3
\nn 
         +224 z^4\big) H_{0,1}(z)
+8 z {\zb}^2 {\zbb} {\zbf}^2 \big(
        16-174 z+661 z^2-988 z^3+395 z^4+334 z^5-376 z^6
\nn 
         +96 z^7\big) \
H_{1,0}(z)
-32 z {\zb}^2 \big(
        8+9 z-20 z^2-13 z^3+15 z^4+3 z^5\big) H_{-1,0}(z)
\nn
 -32 z {\zbb} H_{\frac{1}{2},0,0}(z)
-32 z {\zbb} H_{\frac{1}{2},1,0}(z)
+\frac{4 {\zb}^3 {\zbb}}{3} \big(
        3+91 z-7 z^2-691 z^3+264 z^4+976 z^5
\nn 
         -500 z^6-388 z^7+60 z^8+120 z^9\big) H_{0,0,0}
(z)
+
\frac{16}{3} (1+2 z)^2 \big(
        5-5 z+12 z^2\big) {\zb} H_{0,0,1}(z)
\nn 
 -\frac{8 {\zb}^3 {\zbb}}{3} \big(
        -4+9 z-7 z^2+85 z^3-253 z^4+310 z^5-144 z^6+16 z^7\big) H_{0,1,0}(z)
\nn
 +\frac{64}{3} (1-4 z) (1+2 z)^2 H_{0,1,1}(z)
+\frac{32 {\zb}^3}{3} \big(
        2-2 z-8 z^2-12 z^3+44 z^4-25 z^5-23 z^6
\nn 
         +9 z^7+9 z^8\big) H_{0,-1,0}(z)
-\frac{16 {\zb} {\zbb}}{3} \big(
        -1+5 z-6 z^2-32 z^3+64 z^4+24 z^5\big) H_{1,0,0}(z)
\nn
 +32 z (1+3 z) H_{1,-1,0}(z)
+16 z (1+z)^2 \big(
        -12+11 z+5 z^2+5 z^3\big) {\zb} H_{-1,0,0}(z)
\nn
 -32 z (1+z)^2 (3-4 z) {\zb} H_{-1,1,0}(z)
+32 z (1+z)^2 \big(
        3-z-3 z^2-3 z^3\big) {\zb} H_{-1,-1,0}(z)
\nn
 -192 \text{ln2} z \big(
        -1-z+2 z^2+z^3\big) {\zb} \zeta_2
+\Big(
        -\frac{8}{3} {\zb} {\zbb} {\zbf}^2 {\ztt} \big(
                -14+96 z+170 z^2-2141 z^3
\nn 
                 +735 z^4+15194 z^5-17924 \
z^6-17544 z^7+21248 z^8+1152 z^9\big)
        +224 z {\zbb} H_{\frac{1}{2}}(z)
\nn 
         -8 {\zb} {\zbb} \big(
                2+z-19 z^2+66 z^3-36 z^4-90 z^5+30 z^6+12 z^7\big) H_0(z)
\nn
         -16 z \big(
                11-z+6 z^2\big) {\zbb} H_1(z)
        -16 z (1+z)^2 \big(
                3-7 z+3 z^2+3 z^3\big) {\zb}
         H_{-1}(z)
\Big) \zeta_2
\nn
 +\frac{16}{3} z {\zb}^3 {\zbb} \big(
        14-17 z-209 z^2+546 z^3-241 z^4-385 z^5+352 z^6-18 z^7-36 z^8\big) \zeta_3
%
\nn
 +
48 t {\xxh} {\xxp}^4 H_{-1,0,0}(t) P_2
-8 t (-1+2 t) {\xxm}^4 {\xxp}^3 {\xxt} H_0(t) P_4
+8 t {\xxm}^4 {\xxp}^2 {\xxt} H_{-1}(t) P_4
\nn
 -64 t {\xxm}^5 {\xxp}^5 H_{0,0,-1}(t) P_5
-24 t {\xxm}^5 {\xxp}^5 H_{0,\{3,0\},0}(t) P_5
+48 t {\xxm}^5 {\xxp}^5 H_{0,\{3,0\},-1}(t) P_5
\nn
 -48 t {\xxm}^5 {\xxp}^5 H_{0,\{3,1\},0}(t) P_5
+96 t {\xxm}^5 {\xxp}^5 H_{0,\{3,1\},-1}(t) P_5
-16 t {\xxm}^5 {\xxp}^5 H_{0,-1,0}(t) P_5
\nn 
+32 t {\xxm}^5 {\xxp}^5 H_{0,-1,-1}(t) P_5
-64 t {\xxh} {\xxm}^5 {\xxp}^5 \zeta_3 P_6
+8 t {\xxh} {\xxm}^5 {\xxp}^5 H_{0,0,0}(t) P_9
+12 t {\xxh} {\xxm}^4 {\xxp}^4 {\xxt}^2 H_{0,0}(t) P_{11}
\nn
 +i \pi \Big(
        16 t {\xxh} {\xxp}^4 \zeta_2 P_1
        +32 t {\xxh} {\xxp}^4 H_{-1,0}(t) P_2
        +8 t {\xxm}^3 {\xxp}^3 {\xxt} P_4
        +16 t {\xxh} {\xxm}^5 {\xxp}^5 H_{0,0}(t) P_7
\nn 
         +8 t {\xxh} {\xxm}^4 {\xxp}^4 {\xxt}^2 H_0(t) P_{11}
        -96 t {\xxh} H_{\{3,0\},0}(t)
        -192 t {\xxh} H_{\{3,1\},0}(t)
        +256 t {\xxh} H_{\{4,1\},0}(t)
\Big)
\nn
 -144 t {\xxh} H_{\{3,0\},0,0}(t)
-288 t {\xxh} H_{\{3,1\},0,0}(t)
+384 t {\xxh} H_{\{4,1\},0,0}(t)
-\Big(
         64 t {\xxh} {\xxp}^4 H_{-1}(t) P_3
\nn 
         +8 t {\xxh} {\xxm}^5 {\xxp}^5 H_0(t) P_8
        +16 t {\xxh} {\xxm}^4 {\xxp}^4 {\xxt}^2 P_{10}
\Big) \zeta_2   \,.
\end{align}
The above polynomials $P_i$ are defined as follows: 
\begin{align*}
P_1 =& \, t^4+4 t^3+10 t^2+4 t+1. \\
P_2 =& \, 3 t^4+12 t^3+22 t^2+12 t+3. \\
P_3 =& \, 5 t^4+20 t^3+34 t^2+20 t+5. \\
P_4 =& \, 8 t^6-23 t^5+32 t^4+2 t^3+32 t^2-23 t+8. \\
P_5 =& \, 3 t^8-10 t^6+38 t^4-10 t^2+3. \\
P_6 =& \, 2 t^{10}-5 t^8-4 t^7+24 t^6+4 t^4+4 t^3-2 t^2+1. \\
P_7 =& \, 11 t^{10}-35 t^8-16 t^7+126 t^6-14 t^4+16 t^3+7 t^2+1. \\
P_8 =& \, 35 t^{10}-119 t^8-64 t^7+420 t^6-140 t^4+64 t^3+49 t^2-5. \\
P_9 =& \, 45 t^{10}-133 t^8-48 t^7+490 t^6+70 t^4+48 t^3-7 t^2+15. \\
P_{10} =& \, 4 t^{12}+10 t^{11}+14 t^{10}+12 t^9+57 t^8+198 t^7+274 t^6+198 t^5+57 t^4+12 t^3 \nn
+14 t^2+10 t+4. \\
P_{11} =& \, 5 t^{12}+10 t^{11}+16 t^{10}+4 t^9+59 t^8+194 t^7+288 t^6+194 t^5+59 t^4+4 t^3 \nn
+16 t^2+10 t+5.
\end{align*}
%
The partonic cross-section from $qg$ initiated channels is given by
\begin{align}
\Delta_{qg,Z}^{(1,1)} =& \,  c^{(4)}_q \Big\{
- \frac{{\zb}}{2} \big(
        25+11 z-105 z^2 \! +57 z^3 \! +8 z^4\big)
- \big(
         1+24 z-91 z^2 \! +92 z^3 \! -16 z^4 \! -16 z^5
\nn        
        +4 z^6\big) {\zb}^2 H_0(z)
+\big(
        16-50 z+58 z^2+8 z^3-4 z^4\big) H_1(z)
-2 {\zb}^2 \big(
        3-20 z+52 z^2
\nn        
        -68 z^3 \! +36 z^4\big) H_{\frac{1}{2},0}(z)
-2 {\zb}^2 \big(
        3-20 z+52 z^2 \! -68 z^3 \! +36 z^4\big) H_{\frac{1}{2},1}(z)
-2 {\zb}^2 \big(
        5-58 z
\nn        
        +121 z^2-112 z^3+50 z^4\big) H_{0,0}(z)
-4 {\zb}^2 \big(
        2-13 z+23 z^2-27 z^3+18 z^4\big) H_{0,1}(z)
\nn
-2 \big(
        4-22 z+19 z^2\big) H_{1,0}(z)
-2 (1+z) (3+5 z) H_{-1,0}(z)
+8 (1-2 z)^2 H_{\frac{1}{2},\frac{1}{2},0}(z)
\nn
+8 (1-2 z)^2 H_{\frac{1}{2},\frac{1}{2},1}(z)
+24 (1-2 z)^2 H_{\frac{1}{2},0,0}(z)
+24 (1-2 z)^2 H_{\frac{1}{2},0,1}(z)
\nn
-6 \big(
        1 \! - \! 4 z \! +8 z^2\big) H_{0,0,0}(z)
-2 \big(
        1 \! - \! 4 z \! +8 z^2\big) H_{0,0,1}(z)
+4 \big(
        1-4 z+8 z^2\big) H_{0,-1,0}(z)
\nn
-24 (1-2 z)^2 H_{1,\frac{1}{2},0}(z)
-24 (1-2 z)^2 H_{1,\frac{1}{2},1}(z)
-16 \big(
        5-16 z+16 z^2\big) H_{1,0,0}(z)
\nn
-48 (1-2 z)^2 H_{1,0,1}(z)
-16 \big(
        1-2 z+2 z^2\big) H_{1,1,0}(z)
+16 \big(
        1-2 z+2 z^2\big) H_{1,-1,0}(z)
\nn
+\Big(
        24 \ln (2) \big(
                1-2 z+2 z^2\big)
        +\big(
                -34+200 z-490 z^2+540 z^3-222 z^4\big) {\zb}^2
\nn                
        +2 \big(
                1-4 z+8 z^2\big) H_0(z)
        -48 \big(
                1-3 z+3 z^2\big) H_1(z)
\Big) \zeta_2
+2 \Big(
        19
        -54 z
        +70 z^2
\nn        
        -9 \big(
                1-4 z+8 z^2\big)
\Big) \zeta_3
%
%
+
P_{12} \Big(
-2 w^2 \xyt \big(
        7 H_{0,0}(w)
        +4 H_{0,1}(w)
        -3 H_{0,-\itwo}(w)
\nn        
        +3 H_{0,\mione}(w)
        -4 H_{0,-1}(w)
        -H_{1,-\itwo}(w)
        -H_{-1,-\itwo}(w)
        +H_{1,0}(w)
        +H_{1,\mione}(w)
\nn
        +H_{-1,0}(w)
        +H_{-1,\mione}(w)
\big)
+6 w^2 \xyt \zeta_2
\Big)
+ P_{13} \Big(
2 w^2 \xyb \big(
        7 H_{0,0,0}(w)
        +4 H_{0,0,1}(w)
\nn        
        -3 H_{0,0,-\itwo}(w)
        +3 H_{0,0,\mione}(w)
        -4 H_{0,0,-1}(w)
        -H_{0,1,-\itwo}(w)
        -H_{0,-1,-\itwo}(w)
\nn
        +7 H_{1,0,0}(w)
         \! + \! 4 H_{1,0,1}(w)
         \! - \! 3 H_{1,0,-\itwo}(w)
         \! + \! 3 H_{1,0,\mione}(w)
         \! - \! 4 H_{1,0,-1}(w)
         \! - \! H_{1,1,-\itwo}(w)
\nn
        -H_{1,-1,-\itwo}(w)
        +7 H_{-1,0,0}(w)
        +4 H_{-1,0,1}(w)
        -3 H_{-1,0,-\itwo}(w)
        +3 H_{-1,0,\mione}(w)
\nn        
        -4 H_{-1,0,-1}(w)
        -H_{-1,1,-\itwo}(w)
        -H_{-1,-1,-\itwo}(w)
        +H_{0,1,0}(w)
        +H_{0,1,\mione}(w)
\nn
        +H_{0,-1,0}(w)
        +H_{0,-1,\mione}(w)
        +H_{1,1,0}(w)
        +H_{1,1,\mione}(w)
        +H_{1,-1,0}(w)
        +H_{1,-1,\mione}(w)
\nn
        +H_{-1,1,0}(w)
        +H_{-1,1,\mione}(w)
        +H_{-1,-1,0}(w)
        +H_{-1,-1,\mione}(w)
\big)
\nn
-6 w^2 \xyb \big(
        H_0(w)
        +H_1(w)
        +H_{-1}(w)
\big) \zeta_2
\Big)
%
%
\Big\}
\end{align}
where the polynomials $P_{12}$ and $P_{13}$ are defined as
\begin{align} \label{eq:poly12}
 P_{12} = (-1+6 z-4 z^2+24 z^3)/z^2 ~~
 \text{and} ~~
 P_{13} = (1-4 z+8 z^2)/z^2 \,.
\end{align}
%
%
The partonic cross-section from $qq$ initiated channel, only with double real emission, is also free of any divergences and is given by
\begin{align}
\Delta_{qq,Z}^{(1,1)} =& \, c^{(4)}_q C_F \Big\{
-\frac{4}{3 z^3} (-1+2 z)  I^{(0,1)}_{ell} 
+2 (1-z) \big(
        -103-172 z-83 z^2+4 z^3\big) \zt^2
\nn        
-\frac{4}{3 z^2} \big(13-28 z+44 z^2+16 z^3\big) \ztt  I^{(2,0)}_{ell}
-\frac{4}{3 z^3} \big( -13-65 z+52 z^2+116 z^3\big) \ztt  I^{(3,0)}_{ell}
\nn
-8 \zt^3 \big(
        23+56 z+64 z^2+27 z^3+6 z^4+z^5\big) H_0(z)
+\frac{2 \zb \zt^2 \ztt}{3 z} \big(
        -65-185 z-455 z^2
\nn        
        -551 z^3+376 z^4+1444 z^5+336 z^6-252 z^7+72 z^8\big) H_{0,0}(z)
-\frac{8}{3 z} \big( 13-28 z+44 z^2
\nn
+16 z^3\big) \ztt H_{0,1}(z)
+8 (1-z) \zt^2 \big(
        8+18 z+5 z^2-2 z^3+z^4\big) H_{1,0}(z)
+16 \zt^2 \big(
        7+17 z
\nn        
        +17 z^2+z^3-3 z^4+z^5\big) H_{-1,0}(z)
-96 (1-2 z) H_{-\frac{1}{2},0,0}(z)
+96 (1-2 z) H_{-\frac{1}{2},-1,0}(z)
\nn
-\frac{2}{3 z} \zb \zt^3 \ztt \big(
        -3-82 z-212 z^2-122 z^3+211 z^4+124 z^5+388 z^6+272 z^7\big) H_{0,0,0}(z)
\nn        
-\frac{8}{3 z} (-1+2 z) \big( 5-5 z+12 z^2\big) \zb H_{0,0,1}(z)
-\frac{8}{3 z} \zb \zt^3 \ztt \big( -2+6 z+19 z^2-83 z^3
\nn
-173 z^4-47 z^5+132 z^6+76 z^7\big) H_{0,1,0}(z)
+\frac{32}{3 z} (-1+2 z) (-1+4 z) H_{0,1,1}(z)
\nn
+\frac{16}{3 z} \zb \zt^3 \big( 2-z-2 z^2+18 z^3-22 z^4-35 z^5-8 z^6\big) H_{0,-1,0}(z)
+\frac{8}{3 z} \zb \zt \big(
         1+25 z
\nn        
        -73 z^2+65 z^3-30 z^4\big) H_{1,0,0}(z)
-64 (1-z)^2 \zt H_{1,-1,0}(z)
+16 \big(
         21+45 z-4 z^2
\nn        
        -44 z^3\big) \zt \ztt H_{-1,0,0}(z)
+64 \big(
         1+4 z+z^2-4 z^3\big) \zt \ztt H_{-1,1,0}(z)
-32 \big(
         9+15 z-2 z^2
\nn        
        -12 z^3\big) \zt \ztt H_{-1,-1,0}(z)
+192 \ln (2) \big(
        -1-2 z+z^2+z^3\big) \zt \ztt \zeta_2
\nn
+\Big(
        \frac{8}{3 z} \zt^2 \ztt \big(
                 13+43 z+172 z^2+250 z^3+82 z^4-20 z^5\big)
        +48 (1-2 z) H_{-\frac{1}{2}}(z)
\nn        
        -\frac{8}{z} \zb \zt^3 \ztt \big(
                1+3 z-3 z^2-33 z^3-42 z^4-12 z^5+10 z^6+4 z^7\big) H_0(z)
\nn                
        -32 (1-z)^2 \zt H_1(z)
        -16 \big(
                5-z-6 z^2+4 z^3\big) \zt \ztt H_{-1}(z)
\Big) \zeta_2
\nn
-\frac{16}{3} \zb \zt^3 \ztt \big(
        5+22 z+7 z^2-50 z^3-62 z^4-5 z^5+11 z^6\big) \zeta_3
\Big\}    \,.
\end{align}

\subsection{The partonic coefficients for W}
 
In this Section, we provide the total partonic cross-sections stemming from all the channels with Feynman diagrams where one or two internal $W$ bosons are exchanged.
Like before, the hadronic cross-section receives contributions from several partonic cross-sections convoluted with the PDFs as:
\begin{align}
 \Delta_{W}^{(1,1)} =&
 \sum_{q \in Q,\bar Q} f_q \otimes f_{\bar q} \otimes \Delta_{q\bar{q},W}^{(1,1)}
+\sum_{q \in Q,\bar Q} f_q \otimes f_{g} \otimes \Delta_{qg,W}^{(1,1)}
+\sum_{q \in Q,\bar Q} f_g \otimes f_{q} \otimes \Delta_{qg,W}^{(1,1)}
\nn
+\sum_{q \in Q,\bar Q} f_q \otimes f_{q} \otimes \Delta_{qq,W}^{(1,1)}
\nn
+\sum_{q \in Q,\bar Q} f_q \otimes f_{\bar q} \otimes 
\bigg(
\left(- \delta_s \Delta r^{(1,1)} + \delta g_Z^{(1,1)} \right) \Delta_{q\bar{q}}^{(0,0)} 
+ 2 {\delta s_W^2}^{(1,1)}   \bar{\Delta}_{q\bar{q}}^{(0,0)} 
\bigg)
\nn
+\sum_{q \in Q,\bar Q} f_q \otimes f_{\bar q} \otimes 
\bigg(
\left(- \delta_s \Delta r^{(0,1)} + \delta g_Z^{(0,1)} \right) \Delta_{q\bar{q}}^{(1,0)} 
+ 2 {\delta s_W^2}^{(0,1)}   \bar{\Delta}_{q\bar{q}}^{(1,0)} 
\bigg)
\nn
+\sum_{q \in Q,\bar Q} f_q \otimes f_{g} \otimes 
\bigg(
\left(- \delta_s \Delta r^{(0,1)} + \delta g_Z^{(0,1)} \right) \Delta_{qg}^{(1,0)} 
+ 2 {\delta s_W^2}^{(0,1)}   \bar{\Delta}_{qg}^{(1,0)} 
\bigg) \,,
%
%
\end{align}
where $\Delta r^{(1,1)}$, $\delta g_Z^{(1,1)}$ and ${\delta s_W^2}^{(1,1)}$ ($\Delta r^{(0,1)}$, $\delta g_Z^{(0,1)}$ and ${\delta s_W^2}^{(0,1)}$) are finite renormalisation constants introduced in Section \ref{sec:renormalisation} and evaluated at \oaas (at \oa). The flag $\delta_s$ is introduced to shorten the notation: its value is 1 in the $G_\mu$ input scheme and 0 in the $\alpha(0)$ scheme. 
$\bar{\Delta}_{ij}^{(m,n)}$ are the finite partonic cross sections, defined as the derivative with respect to 
$\sin^2 \theta_W$ of $\Delta_{ij}^{(m,n)}$:
\begin{equation}
  \bar{\Delta}_{ij}^{(m,n)}
  = \frac{-Q_q C_{v,q}}{C_{v,q}^2 + C_{a,q}^2} \Delta_{ij}^{(m,n)} \, , 
 ~~~~~~~(\text{for} ~ (0,0) ~ \text{and} ~ (1,1)) \, .
\end{equation}
%

We present here the results for $u$-type quark only. One can easily obtain the results for $d$-type quarks through appropriate transformations. For example, $\Delta_{d\bar{d},W}^{(1,1)}$ can be obtained from $\Delta_{u\bar{u},W}^{(1,1)}$ through $\{ u\leftrightarrow d \}$. Below we present the NNLO partonic contributions. In these results, as earlier,
only the quark wave function renormalisation has been performed. The $u\bar{u}$ initiated partonic cross-section is given by
\begin{align}
\label{res:uUW}
\Delta_{u\bar{u},W}^{(1,1)} =&  c^2_u C_F \Big\{
\delta (1-z) \Big(
51
-64  {\rm Li}_4 \big( 1/2 \big) 
-\frac{8 \ln^4 (2)}{3}
-43 \zeta_2
+36 \ln (2) \zeta_2
-32 \ln^2 (2) \zeta_2
\nn
+\frac{96}{5} \zeta_2^2
-14 \zeta_3
\Big)
+ \DD_1 \Big(
-44
+32 \zeta_2
\Big)
-(-3+z) (-2+z) \zb
+\big(
        11-23 z+27 z^2
\nn        
        -17 z^3\big) \zb^2 H_0(z)
-22 (1+z) H_1(z)
+2 \big(
        5+15 z+4 z^2+2 z^3\big) H_{0,0}(z)
-2 \big(
         2-8 z
\nn        
        -3 z^2-z^3\big) H_{1,0}(z)
-2 (1+z) \big(
        7+z^2\big) H_{-1,0}(z)
+8 \big(
        1+2 z+5 z^2\big) \zb H_{0,0,0}(z)
\nn        
+4 \big(
        1+2 z+5 z^2\big) \zb H_{0,1,0}(z)
-4 \big(
        1+2 z+5 z^2\big) \zb H_{0,-1,0}(z)
-4 (5-z) H_{1,0,0}(z)
\nn
-4 (3+z) H_{1,1,0}(z)
+8 (1-z) H_{1,-1,0}(z)
-16 (1+z)^2 \zb H_{-1,0,0}(z)
\nn
-8 (1+z)^2 \zb H_{-1,1,0}(z)
+8 (1+z)^2 \zb H_{-1,-1,0}(z)
+\Big(
        2 \big(
                -8+4 z+5 z^2+z^3\big)
\nn                
        +24 \ln (2) \big(
                1+z^2\big) \zb
        +8 (1+z) H_1(z)
        -4 (1+z)^2 \zb H_{-1}(z)
\Big) \zeta_2
+4 (6+z) \zb \zeta_3 
\Big\}
\nn
+
c_u ( c_u - c_d ) C_F \Big\{
\delta (1-z) \Big(
-20
+64   {\rm Li}_4 \big( 1/2 \big)
+\frac{8 \ln^4 (2)}{3}
+\frac{467}{6} \zeta_2
-36 \ln (2) \zeta_2
\nn
+32 \ln^2 (2) \zeta_2
-\frac{278}{5} \zeta_2^2
+\frac{46}{3} \zeta_3
+\frac{67}{2} \sqrt{3} \GI(r_2)
-20 \sqrt{3} \zeta_2 \GI(r_2)
+7 \sqrt{3} \GI(0,r_2)
\nn
-4 \GI (r_2) \GI (0,r_2)
+24 \GI (0, r_2)^2
\Big)
+ \DD_1 \Big(
16
-64 \zeta_2
-40 \sqrt{3} \GI(r_2)
\Big)
+
20
-16 z
\nn
-4 \big(
        -2+z+2 z^2\big) \zb H_0(z)
+8 (1+z) H_1(z)
-2 \big(
        4+16 z+9 z^2+3 z^3\big) H_{0,0}(z)
\nn        
-4 (2-z) z H_{1,0}(z)
-2 (1+z) \big(
        -9+4 z+z^2\big) H_{\{6,0\},0}(z)
-2 (1+z) \big(
         3-8 z
\nn        
        -2 z^2\big) H_{\{6,1\},0}(z)
+2 (1+z) \big(
        7+z^2\big) H_{-1,0}(z)
-4 \big(
        2+3 z+11 z^2\big) \zb H_{0,0,0}(z)
\nn        
-4 \big(
        1+3 z+4 z^2\big) \zb H_{0,1,0}(z)
+4 z H_{0,\{6,0\},0}(z)
-8 z H_{0,\{6,1\},0}(z)
\nn
+4 \big(
        1+2 z+5 z^2\big) \zb H_{0,-1,0}(z)
+28 H_{1,0,0}(z)
+4 H_{1,1,0}(z)
+4 (2+z) H_{1,\{6,0\},0}(z)
\nn
-8 (2+z) H_{1,\{6,1\},0}(z)
-8 (1-z) H_{1,-1,0}(z)
+16 (1+z)^2 \zb H_{-1,0,0}(z)
\nn
+8 (1+z)^2 \zb H_{-1,1,0}(z)
-8 (1+z)^2 \zb H_{-1,-1,0}(z)
+\Big(
        -24 \ln (2) \big(
                1+z^2\big) \zb
\nn                
        +\frac{1}{3} \big(
                63-13 z-70 z^2-14 z^3\big)
        +\frac{8}{3} (1+z) (1+2 z) \zb H_0(z)
        -\frac{8}{3} (7+8 z) H_1(z)
\nn
        +4 (1+z)^2 \zb H_{-1}(z)
\Big) \zeta_2
-\frac{4}{3} \big(
        24+11 z+22 z^2\big) \zb \zeta_3
+\Big(
        2 (1-z) (-3+z) \sqrt{3}
\nn
        +20 z^2 \zb \sqrt{3} H_0(z)
        -20 (1+z) \sqrt{3} H_1(z)
        -2 (1+z) \big(
                1+4 z+z^2\big) \sqrt{3} H_{\{6,0\}}(z)
\nn
        +10 (1+z) \sqrt{3} H_{\{6,1\}}(z)
        +4 z \sqrt{3} H_{0,\{6,0\}}(z)
        +4 (2+z) \sqrt{3} H_{1,\{6,0\}}(z)
\nn        
        -16 \GI(0,r_2)
\Big) \GI(r_2)
-10 (1+z) \sqrt{3} \GI(0,r_2)
%
\nn
+12 \rrt (1-\rho ) (1+\rho ) \Big(
H_{0,0}(\rho ) 
+ H_{0,\{6,0\}}(\rho ) - 2 H_{0,\{6,1\}}(\rho )
\Big)
\nn
+12 \rb^2 \rrt \big(1-\rho +6 \rho ^2-\rho ^3+\rho ^4\big) \Big(
H_{0,0,0}(\rho ) + H_{0,0,\{6,0\}}(\rho ) - 2 H_{0,0,\{6,1\}}(\rho )
\Big)
\nn
-\Big(
        4 \rrt (1-\rho ) (1+\rho )        
        +4 \rb^2 \rrt \big(
                1-\rho +6 \rho ^2-\rho ^3+\rho ^4\big) H_0(\rho )
\Big) \zeta_2
\Big\}
\nn
+ \delta_{u\bar{u},W}^{(1)} + \delta_{u\bar{u},W}^{(2)} \, ,
\end{align}
where $c_{q}$, with $q \in (u,d)$, was defined in Eq.~(\ref{eq:cq}) and where $\delta_{u\bar{u},W}^{(1)}$ and $\delta_{u\bar{u},W}^{(2)}$ denote the contributions from the double-real channel $u \bar{u} \rightarrow d \bar{d} Z$ with one and two internal $W$ boson, respectively. They are given in the following.
\begin{align}
\label{res:uUW1}
\delta_{u\bar{u},W}^{(1)} &= 
c^2_u C_F \Big\{
\big(
        44-65 z+29 z^2\big) \zb
+\big(
        47-61 z+26 z^2\big) \zb^2 H_0(z)
+6 \big(
        7-3 z-z^2
\nn        
        +z^3\big) \zb^2 H_{0,0}(z)
-2 \big(
        -7+7 z-7 z^2+3 z^3\big) \zb^2 H_{1,0}(z)
-4 (1+z) \big(
        7-8 z
\nn        
        +3 z^2\big) \zb^2 H_{-1,0}(z)
+24 \zb^3 \big(
        1-z+3 z^2-3 z^3+z^4\big) H_{0,0,0}(z)
+8 \zb^3 \big(
        1-z+3 z^2
\nn        
        -3 z^3+z^4\big) H_{0,1,0}(z)
-16 \zb^3 \big(
        1-z+3 z^2-3 z^3+z^4\big) H_{0,-1,0}(z)
-12 z \zeta_2
\nn
-8 \zb^3 \big(
        1-z+3 z^2-3 z^3+z^4\big) \zeta_3 
\Big\}
+
c^2_d C_F \Big\{
\frac{1}{2} (1-z) \big(
        -21+27 z+8 z^2\big)
\nn        
-z \big(
        -14+3 z+4 z^2\big) H_0(z)
+2 \big(
        5+28 z+35 z^2+10 z^3\big) H_{0,0}(z)
-4 \big(
        4+18 z+21 z^2
\nn        
        +6 z^3\big) H_{-1,0}(z)
+4 \big(
        3+20 z+30 z^2+20 z^3+5 z^4\big) H_{0,0,0}(z)
-8 \big(
        2+12 z+18 z^2
\nn        
        +12 z^3+3 z^4\big) H_{0,-1,0}(z)
-20 (1+z)^4 H_{-1,0,0}(z)
+24 (1+z)^4 H_{-1,-1,0}(z)
\nn
+\Big(
        -2 \big(
                4+18 z+21 z^2+6 z^3\big)
        -4 \big(
                2+12 z+18 z^2+12 z^3+3 z^4\big) H_0(z)
\nn                
        +12 (1+z)^4 H_{-1}(z)
\Big) \zeta_2
-12 \big(
        1+8 z+12 z^2+8 z^3+2 z^4\big) \zeta_3
\Big\}
\nn
+
c_u c_d C_F \Big\{
\frac{2}{3 z^3} (1+2 z)^2   I^{(0,1)}_{ell}
+\frac{2}{3 z^3}  \ztt \big(
        7+77 z+264 z^2+362 z^3+136 z^4\big) I^{(3,0)}_{ell}
\nn        
+\frac{2}{3 z^2}  \ztt \big(
        -7-32 z+6 z^2+232 z^3+224 z^4\big)  I^{(2,0)}_{ell}        
+6 (1-2 z) (1+z) \zb H_0(z)
\nn
-4 (-7+10 z)
-\frac{1}{3 z} \zb \ztt \big(
        35+163 z+12 z^2-392 z^3+734 z^4+1140 z^5\big) H_{0,0}(z)
\nn        
+\frac{4}{3 z} \ztt \big(
        -7-32 z+6 z^2+232 z^3+224 z^4\big) H_{0,1}(z)
-2 \big(
        5+3 z^2\big) H_{1,0}(z)
\nn        
-4 \big(
        8+28 z +21 z^2\big) H_{-1,0}(z)
-\frac{\zb}{3 z} \big(
        -3-25 z-108 z^2-48 z^3+16 z^4\big) H_{0,0,0}(z)
\nn
+\frac{4}{3 z} (1+2 z)^2 \big( 5-5 z+12 z^2\big) \zb H_{0,0,1}(z)
+\frac{4}{3 z} \zb \big(
        2-z-9 z^2-35 z^3+4 z^4\big) H_{0,1,0}(z)
\nn        
-\frac{16}{3 z} (1+2 z)^2 (-1+4 z) H_{0,1,1}(z)
+\frac{8}{3 z} \zb \big(
        2+11 z+24 z^2+13 z^3+4 z^4\big) H_{0,-1,0}(z)
\nn        
+\frac{4}{3 z} \zb \big(
        1+3 z^2+38 z^3+12 z^4\big) H_{1,0,0}(z)
+8 (1+3 z) H_{1,-1,0}(z)
+16 (1+z)^2 (-3
\nn
+4 z) \zb H_{-1,0,0}(z)
+8 (1+z)^2 (-3+4 z) \zb H_{-1,1,0}
(z)
-8 (1+z)^2 (-3+4 z) \zb H_{-1,-1,0}(z)
\nn
-48 \ln (2) \big(
        -1-z+2 z^2+z^3\big) \zb \zeta_2
+\Big(
        -\frac{2}{3 z} \ztt \big(
                -14-25 z+174 z^2+704 z^3+592 z^4\big)
\nn                
        -\frac{4}{z} \zb \big(
                1+z-9 z^2+14 z^3+24 z^4\big) H_0(z)
        +4 (1+3 z) H_1(z)
\nn        
        +4 (1+z)^2 (-3+4 z) \zb H_{-1}(z)
\Big) \zeta_2
-\frac{4}{3} \big(
        -17-72 z+63 z^2+113 z^3\big) \zb \zeta_3
\Big\}  \,.
\end{align}
\begin{align}
\label{res:uUW2}
\delta_{u\bar{u},W}^{(2)} &= 
c_u ( c_u - c_d ) C_F \Big\{
-\frac{2}{3} \big(
        35
        -11 z
        +30 z \sqrt{3} \GI(0,r_2)
\big)
+\frac{2}{3} \big(
        -33+7 z+2 z^2\big) \zb H_0(z)
\nn        
-\frac{2}{3} \big(
        6+30 z+60 z^2+5 z^3\big) H_{0,0}(z)
-\frac{2}{3} \big(
        2-3 z+12 z^2+z^3\big) H_{1,0}(z)
+2 \big(
        -4+27 z
\nn        
        +12 z^2+z^3\big) H_{\{3,0\},0}(z)
+2 \big(
        -8+9 z+24 z^2+2 z^3\big) H_{\{3,1\},0}(z)
-\frac{4}{3} (1+z) \big(
        -14
\nn        
        +11 z+z^2\big) H_{-1,0}(z)
+4 (1+2 z) (1+5 z) \zb H_{0,0,0}(z)
+4 z (2+z) \zb H_{0,1,0}(z)
\nn
-6 \big(
        1+z+7 z^2\big) \zb H_{0,\{3,0\},0}(z)
-12 \big(
        1+z+7 z^2\big) \zb H_{0,\{3,1\},0}(z)
\nn        
+8 \big(
        1-z+6 z^2\big) \zb H_{0,-1,0}(z)
+\frac{2}{3} (4+15 z) \zeta_2
+\frac{4}{3} \big(
        7+z+28 z^2\big) \zb \zeta_3
\nn
-12 \rrt (1-\rho ) (1+\rho ) \Big(
H_{0,0}(\rho ) + H_{0,\{6,0\}}(\rho ) - 2 H_{0,\{6,1\}}(\rho )
\Big)
\nn
-12 \rb^2 \rrt \big(
        1-\rho +6 \rho ^2-\rho ^3+\rho ^4\big) \Big(
H_{0,0,0}(\rho ) + H_{0,0,\{6,0\}}(\rho ) - 2 H_{0,0,\{6,1\}}(\rho )
\Big)
\nn
+\big(
        4 \rrt (1-\rho ) (1+\rho )
        +4 \rb^2 \rrt \big(
                1-\rho +6 \rho ^2-\rho ^3+\rho ^4\big) H_0(\rho )
\big) \zeta_2
\Big\}
\nn
+
c_d ( c_u - c_d ) C_F \Big\{
2 \big(
        -7
        +7 z
        +10 \sqrt{3} \GI(0,r_2)
        +12 z \sqrt{3} \GI(0,r_2)
\big)
-2 (6+z) H_0(z)
\nn
+2 \big(
        -1+14 z+10 z^2\big) H_{0,0}(z)
-6 \big(
        9+16 z+5 z^2\big) H_{\{3,0\},0}(z)
-6 \big(
        3+14 z
\nn        
        +10 z^2\big) H_{\{3,1\},0}(z)
+4 \big(
        5+14 z+10 z^2\big) H_{-1,0}(z)
+4 \big(
        1+11 z+9 z^2\big) H_{0,0,0}(z)
\nn        
-6 \big(
        1+3 z+3 z^2\big) H_{0,\{3,0\},0}(z)
-12 \big(
        1+3 z+3 z^2\big) H_{0,\{3,1\},0}(z)
-8 (-1+z) H_{0,-1,0}(z)
\nn
-48 (1+z)^2 H_{-1,0,0}(z)
+36 (1+z)^2 H_{-1,\{3,0\},0}(z)
+72 (1+z)^2 H_{-1,\{3,1\},0}(z)
\nn
-24 (1+z)^2 H_{-1,-1,0}(z)
+\Big(
        4
        -4 z (4+3 z) H_0(z)
        +12 (1+z)^2 H_{-1}(z)
\Big) \zeta_2
\nn
+\frac{20}{3} (5+3 z) \zeta_3
\Big\} \,.
\end{align}
We note that once we combine Eq.~\eqref{res:uUW2} with Eq.~\eqref{res:uUW}, the total $\rho$ dependence vanishes.

%
\noindent
The partonic cross-section from $ug$ initiated channels is given by
\begin{align}
\label{res:ugW}
\Delta_{ug,W}^{(1,1)} &=  c^2_u \Big\{
-\frac{\zb}{8} \big(
        1+31 z-73 z^2+29 z^3+8 z^4\big)
-\frac{\zb^2}{4} \big(
        -1+24 z-81 z^2+78 z^3-10 z^4
\nn        
        -16 z^5+4 z^6\big) H_0(z)
+\frac{1}{2} \big(
        5-20 z+22 z^2+4 z^3-2 z^4\big) H_1(z)
+\frac{\zb^2}{2} \big(
        1-21 z+38 z^2
\nn        
        -33 z^3+19 z^4\big) H_{0,0}(z)
+\frac{1}{2} P_{15} \zb^2 H_{0,1}(z)
+\frac{1}{2} \big(
        1+8 z-10 z^2\big) H_{1,0}(z)
+\frac{\zb^2}{2} \big(
        7-39 z
\nn        
        +61 z^2-45 z^3+19 z^4\big) H_{\{6,0\},0}(z)
+\frac{1}{2} P_{15} \zb^2 H_{\{6,0\},1}(z)
-\frac{\zb^2}{2} \big(
        -1-30 z+71 z^2
\nn        
        -72 z^3+38 z^4\big) H_{\{6,1\},0}(z)
-P_{15} \zb^2 H_{\{6,1\},1}(z)
+3 P_{14} H_{0,0,0}(z)
+P_{14} H_{0,0,1}(z)
\nn
+2 P_{14} H_{0,\{6,0\},0}(z)
+P_{14} H_{0,\{6,0\},1}(z)
-4 P_{14} H_{0,\{6,1\},0}(z)
-2 P_{14} H_{0,\{6,1\},1}(z)
\nn
+\big(
        7-18 z+30 z^2\big) H_{1,0,0}(z)
+2 P_{14} H_{1,0,1}(z)
-3 \big(
        1-2 z+2 z^2\big) H_{1,1,0}(z)
\nn
+3 \big(
        2-5 z+8 z^2\big) H_{1,\{6,0\},0}(z)
+2 P_{14} H_{1,\{6,0\},1}(z)
-6 \big(
        2-5 z+8 z^2\big) H_{1,\{6,1\},0}(z)
\nn
-4 P_{14} H_{1,\{6,1\},1}(z)
+\Big(
        \frac{\zb^2}{12} \big(
                45-214 z+381 z^2-260 z^3+38 z^4\big)
        +\frac{2}{3} P_{14} H_0(z)
\nn        
        +\frac{1}{3} \big(
                13-21 z+6 z^2\big) H_1(z)
\Big) \zeta_2
+\big(
        -1+3 z-6 z^2\big) \zeta_3
+\Big(
        -\frac{1}{4} \big(
                -4+32 z-61 z^2
\nn                
                +35 z^3\big) \zb \sqrt{3}
        +z (-2+5 z) \sqrt{3} H_0(z)
        +5 \big(
                1-2 z+2 z^2\big) \sqrt{3} H_1(z)
        +\frac{\zb^2}{2} \big(
                -1-z
\nn                
                +11 z^2-11 z^3+z^4\big) \sqrt{3} H_{\{6,0\}}(z)
        -\frac{1}{2} (-5+6 z) \sqrt{3} H_{\{6,1\}}(z)
        +(2-3 z) \sqrt{3} H_{1,\{6,0\}}(z)
\nn        
        +2 (-2+3 z) \GI(0,r_2)
\Big) \GI(r_2)
+\frac{1}{2} (-5+6 z) \sqrt{3} \GI(0,r_2)
\Big\}
\nn
+ 
c_u c_d \Big\{
\frac{1}{2} (-1+z) (6+7 z)
-\frac{1}{2} \big(
        1+z-4 z^2+3 z^3\big) \zb H_0(z)
+\frac{1}{2} \big(
        3-5 z+7 z^2\big) H_1(z)
\nn        
+\frac{\zb^2}{2} \big(
        -3+20 z-52 z^2+68 z^3-36 z^4\big) H_{\frac{1}{2},0}(z)
+\frac{\zb^2}{2} \big(
        -3+20 z-52 z^2+68 z^3
\nn        
        -36 z^4\big) H_{\frac{1}{2},1}(z)
+\frac{\zb^2}{2} \big(
        -6+79 z-159 z^2+145 z^3-69 z^4\big) H_{0,0}(z)
+\frac{\zb^2}{2} \big(
        -3+29 z
\nn        
        -54 z^2+65 z^3-45 z^4\big) H_{0,1}(z)
-\frac{1}{2} (1-z) (5-9 z) H_{1,0}(z)
+\frac{\zb^2}{2} \big(
        -7+39 z-61 z^2
\nn        
        +45 z^3-19 z^4\big) H_{\{6,0\},0}(z)
+\frac{\zb^2}{2} \big(
        1+3 z-8 z^2+11 z^3-9 z^4\big) H_{\{6,0\},1}(z)
\nn        
+\frac{\zb^2}{2} \big(
        -1-30 z+71 z^2-72 z^3+38 z^4\big) H_{\{6,1\},0}(z)
+P_{15} \zb^2 H_{\{6,1\},1}(z)
\nn
-\frac{1}{2} (1+z) (3+5 z) H_{-1,0}(z)
+2 (1-2 z)^2 H_{\frac{1}{2},\frac{1}{2},0}(z)
+2 (1-2 z)^2 H_{\frac{1}{2},\frac{1}{2},1}(z)
\nn
+6 (1-2 z)^2 H_{\frac{1}{2},0,0}(z)
+6 (1-2 z)^2 H_{\frac{1}{2},0,1}(z)
-\frac{3}{2} \big(
        3-10 z+20 z^2\big) H_{0,0,0}(z)
\nn        
+\frac{1}{2} \big(
        -3+10 z-20 z^2\big) H_{0,0,1}(z)
-2 \big(
        1-3 z+6 z^2\big) H_{0,\{6,0\},0}(z)
\nn
+\big(
        -1+3 z-6 z^2\big) H_{0,\{6,0\},1}(z)
+4 P_{14} H_{0,\{6,1\},0}(z)
+2 \big(
        1-3 z+6 z^2\big) H_{0,\{6,1\},1}(z)
\nn        
+P_{13} z^2 H_{0,-1,0}(z)
-6 (1-2 z)^2 H_{1,
\frac{1}{2},0}(z)
-6 (1-2 z)^2 H_{1,\frac{1}{2},1}(z)
+\big(
        -27+82 z
\nn        
        -94 z^2\big) H_{1,0,0}(z)
-2 \big(
        7-27 z+30 z^2\big) H_{1,0,1}(z)
+\big(
        -1+2 z-2 z^2\big) H_{1,1,0}(z)
\nn        
-3 \big(
        2-5 z+8 z^2\big) H_{1,\{6,0\},0}(z)
-2 \big(
        1-3 z+6 z^2\big) H_{1,\{6,0\},1}(z)
+6 \big(
        2-5 z
\nn        
        +8 z^2\big) H_{1,\{6,1\},0}(z)
+4 P_{14} H_{1,\{6,1\},1}(z)
+4 \big(
        1-2 z+2 z^2\big) H_{1,-1,0}(z)
+6 \ln (2) \big(
        1
\nn        
        -2 z+2 z^2\big) \zeta_2
+\Big(
        \frac{\zb^2}{12} \big(
                -147+814 z-1851 z^2+1880 z^3-704 z^4\big)
        -\frac{1}{6} H_0(z)
\nn        
        +\frac{1}{3} \big(
                -49+129 z-114 z^2\big) H_1(z)
\Big) \zeta_2
+\big(
        6-12 z+5 z^2\big) \zeta_3
+\Big(
        -\frac{1}{4} \big(
                4-32 z+61 z^2
\nn                
                -35 z^3\big) \zb \sqrt{3}
        +(2-5 z) z \sqrt{3} H_0(z)
        -5 \big(
                1-2 z+2 z^2\big) \sqrt{3} H_1(z)
        +\frac{\zb^2}{2} \big(
                1+z-11 z^2
\nn                
                +11 z^3-z^4\big) \sqrt{3} H_{\{6,0\}}(z)
        +\frac{1}{2} (-5+6 z) \sqrt{3} H_{\{6,1\}}(z)
        +(-2+3 z) \sqrt{3} H_{1,\{6,0\}}(z)
\nn        
        -2 (-2+3 z) \GI(0,r_2)
\Big) \GI(r_2)
+\frac{1}{2} (5-6 z) \sqrt{3} \GI(0,r_2)
%
+ P_{12} \Big(
-\frac{w^2 \xyt}{2} \big(
        7 H_{0,0}(w)
\nn        
        +4 H_{0,1}(w)
        +3 H_{0,\mione}(w)
        -3 H_{0,-\itwo}(w)
        -4 H_{0,-1}(w)
        -H_{1,-\itwo}(w)
        -H_{-1,-\itwo}(w)
\nn        
        +H_{1,0}(w)
        +H_{1,\mione}(w)
        +H_{-1,0}(w)
        +H_{-1,\mione}(w)
\big)
+\frac{3}{2} w^2 \xyt \zeta_2
\Big)
+ P_{13} \Big(
\frac{w^2 \xyb}{2} \big(
        7 H_{0,0,0}(w)
\nn        
        +4 H_{0,0,1}(w)
        +3 H_{0,0,\mione}(w)
        -3 H_{0,0,-\itwo}(w)
        -4 H_{0,0,-1}(w)
        -H_{0,1,-\itwo}(w)
\nn        
        -H_{0,-1,-\itwo}(w)
        +7 H_{1,0,0}(w)
        +4 H_{1,0,1}(w)
        +3 H_{1,0,\mione}(w)
        -3 H_{1,0,-\itwo}(w)
\nn        
        -4 H_{1,0,-1}(w)
        -H_{1,1,-\itwo}(w)
        -H_{1,-1,-\itwo}(w)
        +7 H_{-1,0,0}(w)
        +4 H_{-1,0,1}(w)
\nn        
        +3 H_{-1,0,\mione}(w)
        -3 H_{-1,0,-\itwo}(w)
        -4 H_{-1,0,-1}(w)
        -H_{-1,1,-\itwo}(w)
        -H_{-1,-1,-\itwo}(w)
\nn        
        +H_{0,1,0}(w)
        +H_{0,1,\mione}(w)
        +H_{0,-1,0}(w)
        +H_{0,-1,\mione}(w)
        +H_{1,1,0}(w)
        +H_{1,1,\mione}(w)
\nn        
        +H_{1,-1,0}(w)
        +H_{1,-1,\mione}(w)
        +H_{-1,1,0}(w)
        +H_{-1,1,\mione}(w)
        +H_{-1,-1,0}(w)
\nn        
        +H_{-1,-1,\mione}(w)
\big)
-\frac{3}{2} w^2 \xyb \big(
        H_0(w)
        +H_1(w)
        +H_{-1}(w)
\big) \zeta_2
\Big)
\Big\} \,.
\end{align}
$P_{12}$ and $P_{13}$ are given in Eq.~\eqref{eq:poly12} and new polynomials are defined as
\begin{equation}
 P_{14} = (1 - 3 z + 6 z^2) ~~ \text{and} ~~ P_{15} = (-1 - 3 z + 8 z^2 - 11 z^3 + 9 z^4) \,.
\end{equation}

\noindent
The partonic cross-section from $ud$ initiated channel, contributing to the double-real corrections, is also free of any divergences and is given by
%
\begin{align}
\label{res:udW}
\Delta_{ud,W}^{(1,1)} &= \, (c^2_u + c^2_d) C_F \Big\{
-\frac{1}{6 z^3} (-1+2 z)   I^{(0,1)}_{ell}
-\frac{1}{6 z^2} \big( 13-28 z+44 z^2+16 z^3\big) \ztt  I^{(2,0)}_{ell}
\nn
-\frac{1}{6 z^3} \big( -13-65 z+52 z^2+116 z^3\big) \ztt  I^{(3,0)}_{ell}
-\frac{1}{4} (1-z) \big(
        123+200 z+95 z^2\big) \zt^2
\nn        
-\zt^3 \big(
        27+63 z+76 z^2+37 z^3+6 z^4\big) H_0(z)
+\frac{1}{12 z} \zb \zt^2 \ztt \big(
        -65-245 z-287 z^2
\nn        
        +325 z^3+544 z^4+376 z^5+72 z^6\big) H_{0,0}(z)
-\frac{1}{3 z} \big(
        13-28 z+44 z^2+16 z^3\big) \ztt H_{0,1}(z)
\nn        
+(1-z) \big(
        -3+z^2\big) \zt^2 H_{1,0}(z)
+2 (3+z) \zt^2 H_{-1,0}(z)
-\frac{1}{12 z} \zb \zt^3 \big(
        -3+44 z-204 z^2
\nn        
        -2 z^3+167 z^4+150 z^5+40 z^6\big) H_{0,0,0}(z)
-\frac{1}{3 z} (-1+2 z) \big(
        5-5 z+12 z^2\big) \zb H_{0,0,1}(z)
\nn        
-\frac{\zb \zt^3}{3 z} \big(
        -2+22 z-z^2-33 z^3-35 z^4+11 z^5+14 z^6\big) H_{0,1,0}(z)
+\frac{4}{3 z} (-1+2 z) (-1
\nn
              +4 z) H_{0,1,1}(z)
-\frac{2}{3 z} \zb \zt^3 \big(
        -2-5 z+20 z^2+18 z^3+10 z^4+5 z^5+2 z^6\big) H_{0,-1,0}(z)
\nn        
-\frac{1}{3 z} (-1+2 z) \big(
        1+2 z+3 z^2\big) \zb H_{1,0,0}(z)
+\Big(
        \frac{1}{3 z} \zt^2 \ztt \big(
                13-2 z+13 z^2+103 z^3+79 z^4
\nn                
                +10 z^5\big)
        +\frac{\zb \zt^3}{z} \big(
                -1-3 z-z^2-5 z^3+8 z^4+18 z^5+8 z^6\big) H_0(z)
\Big) \zeta_2
\nn
+\frac{2}{3} \zb \zt^3 \big(
        -8-18 z-19 z^2+19 z^3+36 z^4+14 z^5\big) \zeta_3
\Big\} 
\nn
+
c_u c_d C_F \Big\{
2 (1-z) \big(
        5+2 z+z^2\big) \zt
-2 \zt^2 \big(
        -4-3 z-9 z^2-z^3+z^4\big) H_0(z)
\nn
-2 \zt \big(
        -5+24 z+20 z^2-12 z^3+3 z^4\big) H_{0,0}(z)
+2 (1-z) \big(
        11-4 z+z^2\big) H_{1,0}(z)
\nn        
+4 \zt \big(
        4+12 z+5 z^2-4 z^3+z^4\big) H_{-1,0}(z)
-24 (1-2 z) H_{-\frac{1}{2},0,0}(z)
\nn
+24 (1-2 z) H_{-\frac{1}{2},-1,0}(z)
+4 \big(
        5-z-6 z^2+8 z^3\big) \zt \ztt H_{0,0,0}(z)
+8 \big(
        1+4 z^2\big) \ztt H_{0,1,0}(z)
\nn        
+8 \big(
        -1+4 z+z^2\big) \zt H_{0,-1,0}(z)
+16 (1-z)^2 \zt H_{1,0,0}(z)
-16 (1-z)^2 \zt H_{1,-1,0}(z)
\nn
-4 \big(
        -21-45 z+4 z^2+44 z^3\big) \zt \ztt H_{-1,0,0}(z)
-16 \big(
        -1-4 z-z^2+4 z^3\big) \zt \ztt H_{-1,1,0}(z)
\nn        
+8 \big(
        -9-15 z+2 z^2+12 z^3\big) \zt \ztt H_{-1,-1,0}(z)
+48 \ln (2)  \big(
        -1-2 z+z^2+z^3\big) \zt \ztt \zeta_2
\nn
+\Big(
        -2 \big(
                -15-8 z+5 z^2\big) \zt
        +12 (1-2 z) H_{-\frac{1}{2}}(z)
        +4 \big(
                1+4 z+17 z^2+10 z^3\big) \zt \ztt H_0(z)
\nn                
        -8 (1-z)^2 \zt H_1(z)
        -4 \big(
                5-z-6 z^2+4 z^3\big) \zt \ztt H_{-1}(z)
\Big) \zeta_2
\nn
+4 \big(
        1+3 z+14 z^2+13 z^3\big) \zt \ztt \zeta_3
\Big\}  
+ \delta_{ud,W} \,.
%
\\
\delta_{ud,W} &= \, ( c_u - c_d )^2 C_F \Big\{
\frac{1}{3} (1-z) \big(
        53+44 z+3 z^2\big) \zt
\nn        
-\frac{\zt^2}{3} \big(
        -30-77 z-85 z^2-11 z^3+3 z^4\big) H_0(z)
\nn        
+\frac{1}{3} \big(
        15+27 z-90 z^2+11 z^3\big) H_{0,0}(z)
-\frac{1}{3} (1-z) \big(
        -2-11 z+z^2\big) H_{1,0}(z)
\nn        
+\big(
        -29+42 z+27 z^2-4 z^3\big) H_{\{3,0\},0}(z)
+\big(
        -13-6 z+54 z^2-8 z^3\big) H_{\{3,1\},0}(z)
\nn        
+\frac{2}{3} \big(
        13-42 z^2+7 z^3\big) H_{-1,0}(z)
+6 (1-2 z) (1-z) \zt H_{0,0,0}(z)
+2 \big(
        1-z
\nn        
        +4 z^2\big) \zt H_{0,1,0}(z)
-6 (1-2 z) \zt H_{0,\{3,0\},0}(z)
-12 (1-2 z) \zt H_{0,\{3,1\},0}(z)
-4 \big(
        -2+2 z
\nn        
        +z^2\big) \zt H_{0,-1,0}(z)
+8 \big(
        -2-z+4 z^2\big) \zt H_{-1,0,0}(z)
-6 \big(
        -2-z+4 z^2\big) \zt H_{-1,\{3,0\},0}(z)
\nn        
-12 \big(
        -2-z+4 z^2\big) \zt H_{-1,\{3,1\},0}(z)
+4 \big(
        -2-z+4 z^2\big) \zt H_{-1,-1,0}(z)
+\Big(
        \frac{1}{3} (2+3 z)
\nn        
        +2 \zt \big(
                3 z^2
                +\frac{1}{\zt}
        \big) H_0(z)
        -2 \big(
                -2-z+4 z^2\big) \zt H_{-1}(z)
\Big) \zeta_2
-\frac{2}{3} \big(
        -24-16 z+17 z^2\big) \zt \zeta_3
\nn        
+10 (1-2 z) \sqrt{3}
 \GI(0,r_2)
\Big\}  \,.
\end{align}
The contribution $\delta_{ud,W}$ in Eq.~\eqref{res:udW} stems solely from the non-abelian $WWZ$ vertex. The partonic cross-section from $u\bar{d}$ initiated channel is also only with double real emission and given by
\begin{align}
\label{res:uDW}
\Delta_{u\bar{d},W}^{(1,1)} &= \, (c^2_u + c^2_d) C_F \Big\{
-\big(
        -39+323 z-682 z^2+74 z^3\big) \zbf^2
-\zb \zbf^2 \big(
        -26+187 z-265 z^2
\nn        
        -296 z^3+76 z^4\big) H_0(z)
+\big(
        11-78 z+129 z^2+46 z^3\big) \zbf^2 H_{0,0}(z)
+\big(
        11-78 z+129 z^2
\nn        
        +46 z^3\big) \zbf^2 H_{1,0}(z)
+3 (1+3 z) \zb H_{0,0,0}(z)
+3 (1+3 z) \zb H_{0,1,0}(z)
+\Big(
        \big(
                11-78 z+129 z^2
\nn                
                +46 z^3\big) \zbf^2
        +3 (1+3 z) \zb H_0(z)
\Big) \zeta_2
+6 (1+3 z) \zb \zeta_3
%
+ \frac{1}{z} \Big(
-\frac{1}{2} t (-1+2 t) \xxm^4 \xxp^3 H_0(t) P_{16}
\nn
+\frac{1}{2} t \xxm^4 \xxp^2 H_{-1}(t) P_{16}
-2 t \xxm^4 \xxp^4 H_{0,-1}(t) P_{17}
-2 t \xxm^4 \xxp^4 H_{-1,0}(t) P_{17}
+4 t \xxm^4 \xxp^4 H_{-1,-1}(t) P_{17}
\nn
+t \xxm^4 \xxp^4 H_{0,0}(t) P_{19}
-8 t \xxm^5 \xxp^5 H_{0,0,-1}(t) P_{20}
-3 t \xxm^5 \xxp^5 H_{0,\{3,0\},0}(t) P_{20}
\nn
+6 t \xxm^5 \xxp^5 H_{0,\{3,0\},-1}(t) P_{20}
-6 t \xxm^5 \xxp^5 H_{0,\{3,1\},0}(t) P_{20}
+12 t \xxm^5 \xxp^5 H_{0,\{3,1\},-1}(t) P_{20}
\nn
-2 t \xxm^5 \xxp^5 H_{0,-1,0}(t) P_{20}
+4 t \xxm^5 \xxp^5 H_{0,-1,-1}(t) P_{20}
-2 t \xxm^5 \xxp^5 \zeta_3 P_{21}
+t \xxm^5 \xxp^5 H_{0,0,0}(t) P_{23}
\nn
+\frac{1}{2} t \big(
        1+t+t^2\big) \xxm^4 \xxp^4 P_{16}
+i \pi \Big(
        \frac{1}{2} t \xxm^3 \xxp^3 P_{16}
        +2 t \xxm^4 \xxp^4 H_0(t) P_{18}
        +2 t \xxm^5 \xxp^5 H_{0,0}(t) P_{21}
\nn        
        +4 t \big(
                1+t^2\big) \xxp^4 H_{-1,0}(t)
        +2 t \big(
                1+t^2\big) \xxp^4 \zeta_2
\Big)
+6 t \big(
        1+t^2\big) \xxp^4 H_{-1,0,0}(t)
+\big(
        -4 t \xxm^4 \xxp^4 P_{18}
\nn        
        -t \xxm^5 \xxp^5 H_0(t) P_{22}
        -8 t \big(
                1+t^2\big) \xxp^4 H_{-1}(t)
\big) \zeta_2
\Big)
\Big\}
%
\nn
+ c_u c_d C_F \Big\{ 
-\big(
        15-127 z+236 z^2+164 z^3\big) \zbf^2
-2 \zb \zbf^2 \big(
        7-46 z+42 z^2+104 z^3
\nn        
        +28 z^4\big) H_0(z)
+2 \zb \zbb \zbf^2 \big(
        -2+12 z-12 z^2+31 z^3-256 z^4+380 z^5\big) H_{0,0}(z)
\nn        
+2 \zbb \zbf^2 \big(
        1-9 z+z^2+132 z^3-188 z^4\big) H_{1,0}(z)
+12 (1+z)^2 \zb H_{-1,0}(z)
+16 \zbb H_{0,0,0}(z)
\nn
+16 \zbb H_{0,0,1}(z)
-4 \big(
        -2+z+3 z^2+2 z^3\big) \zb \zbb H_{0,1,0}(z)
+8 (1+z)^2 \zb H_{0,-1,0}(z)
\nn
+20 \zbb H_{1,0,0}(z)
+16 \zbb H_{1,0,1}(z)
+8 \zbb H_{1,1,0}(z)
+20 (1+z)^2 \zb H_{-1,0,0}(z)
\nn
-24 (1+z)^2 \zb H_{-1,-1,0}(z)
+\Big(
        2 \zb \zbb \zbf^2 \big(
                4-34 z+49 z^2+197 z^3-416 z^4+92 z^5\big)
\nn                
        -4 \big(
                1-3 z+6 z^2+4 z^3\big) \zb \zbb H_0(z)
        -8 \zbb H_1(z)
        -12 (1+z)^2 \zb H_{-1}(z)
\Big) \zeta_2
-4 \big(
        -1-z
\nn        
        +6 z^2+4 z^3\big) \zb \zbb \zeta_3
%
%
+ \frac{1}{z} \Big(
4 t \xxh \xxp^4 H_{1,0}(z) P_{25}
-32 t \xxh \xxm^5 \xxp^5 H_{0,0,-1}(t) P_{27}
\nn
-4 t \xxh \xxm^5 \xxp^5 H_{0,\{3,0\},0}(t) P_{28}
+8 t \xxh \xxm^5 \xxp^5 H_{0,\{3,0\},-1}(t) P_{28}
-8 t \xxh \xxm^5 \xxp^5 H_{0,\{3,1\},0}(t) P_{28}
\nn
+16 t \xxh \xxm^5 \xxp^5 H_{0,\{3,1\},-1}(t) P_{28}
-4 t \xxh \xxm^5 \xxp^5 \zeta_3 P_{29}
-8 t \xxh \xxm^5 \xxp^5 H_{0,-1,0}(t) P_{30}
\nn
+16 t \xxh \xxm^5 \xxp^5 H_{0,-1,-1}(t) P_{30}
+8 t \xxh \xxm^5 \xxp^5 H_{0,0,0}(t) P_{31}
+2 t \xxh \xxm^4 \xxp^4 \xxt^2 H_{0,0}(t) P_{34}
\nn
-2 t \xxh \xxm^4 \xxp^4 \xxt^2 H_{0,-1}(t) P_{36}
-2 t \xxh \xxm^4 \xxp^4 \xxt^2 H_{-1,0}(t) P_{36}
+4 t \xxh \xxm^4 \xxp^4 \xxt^2 H_{-1,-1}(t) P_{36}
\nn
-3 t \big(
        1+4 t+t^2\big) \xxm^4 \xxp^4 P_{24}
+ i \pi  \Big(
        16 t^2 \xxh \xxm^5 \xxp^5 H_{0,0}(t) P_{26}
        -2 t \xxh \xxm^4 \xxp^4 \xxt^2 H_0(t) P_{33}
\nn        
        -3 t \big(
                1+4 t+t^2\big) \xxm^3 \xxp^3 \xxt P_{24}
        -24 t \xxh H_{\{3,0\},0}(t)
        -48 t \xxh H_{\{3,1\},0}(t)
        +64 t \xxh H_{\{4,1\},0}(t)
\nn        
        +16 t \big(
                1+4 t+t^2\big) \xxh \xxp^2 H_{-1,0}(t)
        +16 t^2 \xxh \xxp^2 \zeta_2
\Big)
+3 t (-1+2 t) \big(
        1+4 t+t^2\big) \xxm^4 \xxp^3 \xxt P_{24} H_0(t)
\nn        
-3 t \big(
        1+4 t+t^2\big) \xxm^4 \xxp^2 \xxt P_{24} H_{-1}(t)
-16 t \xxh H_{0,0,1}(z)
-8 t \xxh H_{0,1,0}(z)
-24 t \xxh H_{1,0,0}(z)
\nn
-16 t \xxh H_{1,0,1}(z)
-8 t \xxh H_{1,1,0}(z)
-36 t \xxh H_{\{3,0\},0,0}(t)
-72 t \xxh H_{\{3,1\},0,0}(t)
\nn
+112 t \xxh H_{\{4,1\},0,0}(t)
-32 t \xxh H_{\{4,1\},0,-1}(t)
-16 t \xxh H_{\{4,1\},\{3,0\},0}(t)
+32 t \xxh H_{\{4,1\},\{3,0\},-1}(t)
\nn
-32 t \xxh H_{\{4,1\},\{3,1\},0}(t)
+64 t \xxh H_{\{4,1\},\{3,1\},-1}(t)
+24 t \big(
        3+8 t+3 t^2\big) \xxh \xxp^2 H_{-1,0,0}(t)
\nn        
-96 t \xxh H_{-1,0,-1}(t)
-96 t \xxh H_{-1,-1,0}(t)
+192 t \xxh H_{-1,-1,-1}(t)
+\Big(
        -4 t \xxh \xxm^5 \xxp^5 H_0(t) P_{32}
\nn        
        +4 t \xxh \xxm^4 \xxp^4 \xxt^2 P_{35}
        -40 t \xxh H_1(z)
        -112 t \xxh H_{\{4,1\}}(t)        
        +16 t \big(
                3+2 t+3 t^2\big) \xxh \xxp^2 H_{-1}(t)
\Big) \zeta_2
\Big) 
\Big\} 
\nn
+ \delta_{u\bar{d},W}\,.
\end{align}
The polynomials here are defined as
\begin{align*}
 P_{16} =& \, 19 t^4-56 t^3+86 t^2-56 t+19.
 \\
 P_{17} =& \, 2 t^6-2 t^5+t^4+4 t^3+t^2-2 t+2.
 \\
 P_{18} =& \, 3 t^6-2 t^5+3 t^4+4 t^3+3 t^2-2 t+3.
 \\
 P_{19} =& \, 11 t^6-8 t^5+10 t^4+16 t^3+10 t^2-8 t+11.
 \\
 P_{20} =& \, t^8+2 t^6+6 t^4+2 t^2+1.
 \\
 P_{21} =& \, 3 t^8-4 t^7+10 t^6-4 t^5+12 t^4+4 t^3-2 t^2+4 t+1.
 \\
 P_{22} =& \, 9 t^8-16 t^7+34 t^6-16 t^5+30 t^4+16 t^3-14 t^2+16 t+1.
 \\
 P_{23} =& \, 13 t^8-12 t^7+38 t^6-12 t^5+60 t^4+12 t^3+2 t^2+12 t+7.
 \\
 P_{24} =& \, t^4-t^3-2 t^2-t+1.
 \\
 P_{25} =& \, t^4+2 t^3+5 t^2+2 t+1.
 \\
 P_{26} =& \, 2 t^9+t^8-12 t^7-2 t^6+26 t^5-6 t^3+2 t^2+2 t-1.
 \\
 P_{27} =& \, 3 t^{10}-15 t^8+30 t^6-10 t^4+5 t^2-1.
 \\
 P_{28} =& \, 5 t^{10}-25 t^8+50 t^6+10 t^4-5 t^2+1.
 \\
 P_{29} =& \, 5 t^{10}+4 t^9-33 t^8-8 t^7+74 t^6+6 t^4+8 t^3-7 t^2-4 t+3.
 \\
 P_{30} =& \, 7 t^{10}-35 t^8+70 t^6-50 t^4+25 t^2-5.
 \\
 P_{31} =& \, 12 t^{10}+3 t^9-66 t^8-6 t^7+138 t^6-38 t^4+6 t^3+16 t^2-3 t-2.
 \\
 P_{32} =& \, 13 t^{10}+8 t^9-81 t^8-16 t^7+178 t^6-78 t^4+16 t^3+31 t^2-8 t-3.
 \\
 P_{33} =& \, t^{12}-2 t^{11}+6 t^{10}+24 t^9-5 t^8-134 t^7-212 t^6-134 t^5-5 t^4+24 t^3 \nn
 +6 t^2-2 t+1.
 \\
 P_{34} =& \, t^{12}+2 t^{11}-10 t^{10}-56 t^9+26 t^8+222 t^7+386 t^6+222 t^5+26 t^4-56 t^3 \nn
 -10 t^2+2 t+1.
 \\
 P_{35} =& \, 2 t^{12}-2 t^{11}+8 t^{10}+16 t^9-3 t^8-138 t^7-198 t^6-138 t^5-3 t^4+16 t^3 \nn
 +8 t^2-2 t+2.
 \\
 P_{36} =& \, 5 t^{12}-2 t^{11}-2 t^{10}-40 t^9+37 t^8+42 t^7+136 t^6+42 t^5+37 t^4-40 t^3 \nn
 -2 t^2-2 t+5.
\end{align*}
%
$\delta_{u\bar{d},W}$ also stems solely from the non-abelian $WWZ$ vertex and is given by
\begin{align}
\delta_{u\bar{d},W} &= \, (c_u-c_d)^2 C_F \Big\{
-6
-\frac{13 z}{2}
+i \pi \Big(
        -\frac{1}{2} (1-t) \xxp (3+4 z) \zb
        -\zb H_0(t)
        -4 z (2+z) \zb H_{0,0}(t)
\nn        
        +8 z (2+z) \zb H_{-1,0}(t)
        +4 z (2+z) \zb \zeta_2
\Big)
+\frac{1}{2} (-1+2 t) \xxp (3+4 z) \zb H_0(t)
\nn
-\frac{1}{2} (1+z) (4+5 z) \zb H_0(z)
-\frac{1}{2} (3+4 z) \zb H_{-1}(t)
-\zb H_{0,0}(t)
-\zb H_{0,-1}(t)
-\zb H_{-1,0}(t)
\nn
+2 \zb H_{-1,-1}(t)
-6 z (2+z) \zb H_{0,0,0}(t)
+12 z (2+z) \zb H_{-1,0,0}(t)
\nn
+\Big(
        2 \zb
        +8 z (2+z) \zb H_0(t)
        -16 z (2+z) \zb H_{-1}(t)
\Big) \zeta_2
+4 z (2+z) \zb \zeta_3
\Big\} \,.
\end{align}

%% file: pheno.tex
\section{Phenomenology}
\label{se:phenomenology}
In this section we present the numerical results for the inclusive total cross section for the production of an on-shell $Z$ boson in  proton-antiproton collisions at the Tevatron, with a collider center-of-mass energy $\sqrt{S}=1.96$ TeV, and in proton-proton collisions at the LHC, with different energies ($\sqrt{S}=7,8,13,14,100$ TeV). They are computed using the following values of the input parameters:
$m_H=125.0\, {\rm GeV}$,
$\mw=80.358\, {\rm GeV}$, 
$\mz=91.153\, {\rm GeV}$, 
$m_t=173.2\, {\rm GeV}$, 
$\alpha^{-1}=137.035999074$,
$G_\mu=1.1663781\,10^{-5}\, {\rm GeV}^{-2}$
and $\Delta\alpha_{had}(\mz)=0.027572$,
where $m_t$ and $m_H$ are the top quark and Higgs boson masses.
We consider 5 active flavours in the proton. In the $b\bar b$ channel, at NLO-EW, we include the exact dependence on $m_t$.
At NNLO QCD-EW, in the 2-loop virtual subset,
we instead use the same amplitude computed for a generic massless quark doublet,
for the $d\bar d$-initiated channel. We consider this approximation acceptable from a phenomenological point of view,
based on the estimate done at NLO-EW for the size of the $m_t$ effect, compared to the massless limit.
We choose different parametrisations of the proton structure and for each set of parton density functions (PDFs) we consider two variants: one determination uses only QCD matrix elements and DGLAP QCD evolution, while the other relies on the same data set but adopts QCD+EW matrix elements and DGLAP QCD+QED evolution. We considered three pairs of PDF sets:
{\tt NNPDF31\_nnlo\_as\_0118} and
{\tt NNPDF31\_nnlo\_as\_0118\_luxqed} \cite{NNPDF:2017mvq},
{\tt MMHT2015\_nnlo} and
{\tt MMHT2015qed\_nnlo} \cite{Harland-Lang:2019pla}, and
{\tt CT18NNLO} \cite{Hou:2019efy} and
{\tt CT18qed} \cite{Xie:2021equ}.
\begin{table}
\begin{center}
\begin{tabular}{|c|c|c|c|c|c|c|}
\hline
  collider & PDF set &  $\sigma_{QCD}$  &  $\sigma_{QCD-EW}$ & $\delta_{QCD-EW} $ & $\Delta_{env} $ & $\delta_{PDF}$\\
\hline\hline
$p\bar p$ 1.96 TeV &  {\tt NNPDF3.1} & 7710.0 & 7649.5 & -0.8 & 0.3\% & $^{+1.7\%}_{-1.7\%}$ \\
\hline
           &  {\tt CT18}     & 7683.8 & 7640.7 & -0.6 & & $^{+1.6\%}_{-2.3\%}$\\
\hline
           &  {\tt MMHT2015} & 7701.1 & 7625.8 & -1.0 & & $^{+2.4\%}_{-2.4\%}$\\
\hline
LHC 7 TeV &  {\tt NNPDF3.1} & 29356.2 & 29120.4& -0.8 &  1.8\% & $^{+0.9\%}_{-0.9\%}$\\
\hline
           &  {\tt CT18}     & 28836.9 & 28702.4 & -0.5 &  & $^{+1.5\%}_{-2.4\%}$ \\
\hline
           &  {\tt MMHT2015} & 29023.0 & 28709.1 & -1.1  &  & $^{+2.0\%}_{-2.1\%}$\\
\hline
LHC 8 TeV &  {\tt NNPDF3.1} & 34116.0 & 33840.2 & -0.8 & 1.6\% & $^{+0.8\%}_{-0.8\%}$\\
\hline
           &  {\tt CT18}     & 33562.2 & 33407.5 & -0.5 & & $^{+1.6\%}_{-2.4\%}$\\
\hline
           &  {\tt MMHT2015} & 33792.4 & 33420.8 & -1.1  &  & $^{+2.0\%}_{-2.1\%}$\\
\hline
LHC 13 TeV &  {\tt NNPDF3.1} & 57769.1 & 57287.6 & -0.8 & 1.1\% & $^{+0.8\%}_{-0.8\%}$ \\
\hline
           &  {\tt CT18}     & 57152.1 & 56898.9 & -0.4 &  & $^{+1.9\%}_{-2.5\%}$\\
\hline
           &  {\tt MMHT2015} & 57564.8 & 56899.3 & -1.2  &  &  $^{+2.1\%}_{-2.1\%}$\\
\hline
LHC 14 TeV &  {\tt NNPDF3.1} & 62454.4 & 61931.2 & -0.8 & 1.0\% &  $^{+0.8\%}_{-0.8\%}$\\
\hline
           &  {\tt CT18}     & 61840.8 & 61568.1 & -0.4 &  &  $^{+2.0\%}_{-2.5\%}$\\
\hline
           &  {\tt MMHT2015} & 62278.6 & 61553.7 & -1.2 &  &  $^{+2.2\%}_{-2.2\%}$\\
\hline
LHC 100 TeV &  {\tt NNPDF3.1} & 418617 & 412815 & -1.4 &  2.4\% & $^{+3.1\%}_{-3.1\%}$ \\
\hline
           &  {\tt CT18}     & 420218 & 418344 & -0.4 &  & $^{+5.5\%}_{-3.8\%}$\\
\hline
           &  {\tt MMHT2015} & 410367 & 405238 & -1.2 &  & $^{+6.4\%}_{-4.4\%}$\\
\hline
\end{tabular}
\caption{\label{tab:pdfresults}
Cross sections for on-shell $Z$ production, expressed in picobarns and computed with different PDF sets at different collider types and energies. The two columns show the results obtained with PDF parameterisations determined with a QCD-only analysis or including also EW effects. We define $\delta_{QCD-EW}=100\, (\sigma_{QCD+EW}/\sigma_{QCD} -1)$, while $\Delta_{env}$ is the percentage width of the envelope of the three PDF sets predictions in the QCD model, with respect to their mean value. The experimental PDF uncertainty $\delta_{PDF}$ in the QCD model is computed according to the definitions of each group.
} 
\end{center}
\end{table}
\begin{table}
\begin{center}
\begin{tabular}{|c|c|c|c|c|}
\hline
  collider &  $\sigma_{QCD}$  & $\delta_{7pts}$ & $\sigma_{QCD-EW}$ & $\delta_{7pts}$ \\
\hline\hline
$p\bar p$ 1.96 TeV & 7710.0 & $^{+0.48\%}_{-0.66\%}$ & 7649.5 & $^{+0.37\%}_{-0.64\%}$  \\
\hline
LHC 7 TeV &  29356.2        & $^{+0.52\%}_{-0.26\%}$ & 29120.4& $^{+0.49\%}_{-0.31\%}$     \\
\hline
LHC 8 TeV &  34116.0        & $^{+0.58\%}_{-0.30\%}$ & 33840.2& $^{+0.56\%}_{-0.35\%}$    \\
\hline
LHC 13 TeV &   57769.1      & $^{+0.78\%}_{-0.45\%}$ & 57287.6& $^{+0.76\%}_{-0.49\%}$     \\
\hline
LHC 14 TeV &  62454.4       & $^{+0.80\%}_{-0.47\%}$ & 61931.2& $^{+0.79\%}_{-0.50\%}$     \\
\hline
LHC 100 TeV &  418617       & $^{+1.26\%}_{-1.16\%}$ & 412815 & $^{+1.26\%}_{-1.18\%}$  \\
\hline
\end{tabular}
\caption{\label{tab:qcdscales}
  Dependence of the cross sections for on-shell $Z$ production, in picobarns, on the renormalisation and factorisation scale choices.
  The upper and lower percentage variations, compared to the central scales choice, is computed among 7 scales combinations (cfr. text).
  The results have been computed with the central replica of the
  {\tt NNPDF31\_nnlo\_as\_0118} and {\tt NNPDF31\_nnlo\_as\_0118\_luxqed} PDF sets.
  } 
\end{center}
\end{table}
We have used the packages 
{\texttt{GiNaC}} \cite{Vollinga:2004sn, Bauer:2000cp},
{\texttt{handyG}} \cite{Naterop:2019xaf} and
{\texttt{HarmonicSums}} \cite{Ablinger:2010kw,Ablinger:2013cf,Ablinger:2014rba},
for manipulation and numerical evaluation of all the polylogarithmic functions which appear in the final expressions of our results.

To present the results, we consider two different approximations of the total cross section:
one that includes only QCD radiative corrections $\sigma_{QCD} \equiv \sigma^{(0,0)} + \alpha_s \sigma^{(1,0)} + \alpha_s^2 \sigma^{(2,0)}$
and one that represents our complete prediction with EW and mixed QCD-EW corrections
$\sigma_{QCD-EW} \equiv \sigma^{(0,0)} + \alpha_s \sigma^{(1,0)} + \alpha \sigma^{(0,1)} + \alpha\alpha_s \sigma^{(1,1)} + \alpha_s^2\sigma^{(2,0)}$
where $\sigma^{(i,j)}$ indicates the sole contribution from the relative perturbative order ${\cal O}(\alpha_s^i \alpha^j)$
with respect to the Born.
In the $\sigma_{QCD}$ prediction the PDFs encode the effect of the EW corrections in their parameterisation,
while their perturbative evolution is driven only by the QCD kernel.
On the other hand, in $\sigma_{QCD+EW}$ the PDFs should be extracted including NLO-EW corrections in the matrix elements
used to fit the data\footnote{A complete systematic analysis including all the NLO-EW corrections is in progress.}
and are evolved with QCD+QED DGLAP kernels.
The main component available in all the considered sets is the photon density, which fulfils the constraints imposed
by the so called LUX-qed model.
In the latter, the photon density is connected by an exact relation to the hadronic tensor
and the proton structure functions \cite{Manohar:2016nzj}.

The two models, the one where only QCD radiative corrections and the one with a complete QCD and EW analysis,
consistently yield a possible description of the proton-proton scattering.
For this reason, we then consider the two predictions $\sigma_{QCD}$ and $\sigma_{QCD+EW}$
as possible alternatives for the best predictions of the on-shell $Z$ production cross section,
a classical benchmark used to compare different proton parameterisations.
The $\sigma_{QCD+EW}$ is preferred, for its richer perturbative content.
The difference between the two cross sections can thus be taken as an estimate of the impact of the mixed QCD-EW corrections
on this observable\footnote{ The inclusion in the future of N3LO results with the (today still missing)
  appropriate N3LO PDFs might partially interfere with our discussion,
  inducing a different mixture of the various partonic channels.}.
We observe in Table \ref{tab:pdfresults} that different PDF choices lead to a spread in the central value prediction;
comparing the width of their envelope with the mean value of the three predictions,
in the QCD-only case,
we observe a dispersion of the results $\Delta_{env}$ ranging between the ${\cal O}(1\%)$ and the ${\cal O}(2\%)$ level at the LHC energies,
while a smaller value is observed at the Tevatron.
This spread is compatible with the estimate of the PDF uncertainty evaluated
according to the Monte Carlo or Hessian recipes used by each PDF collaboration.
We remark that the PDF uncertainties range between the ${\cal O}(1\%)$ and the ${\cal O}(\pm 2.5\%)$ level,
depending on the energy, collider and PDF fitting group; larger values are instead obtained for a collider with $\sqrt{S}=100$ TeV.

For each PDF set, the shift induced by the inclusion of the NLO-EW and NNLO QCD-EW corrections
is almost independent of collider and energy and ranges between -0.4\% and -1.4\%.
The size of this shift of the central value is significant if compared with the residual subpercent QCD scale uncertainty
reported in Refs.\cite{Duhr:2020seh,Duhr:2020sdp} at N3LO-QCD, but also with the corresponding estimate made at NNLO-QCD.
In Table \ref{tab:qcdscales} we present the width of the QCD scale uncertainty band,
evaluated considering independent variations of the renormalisation scale $\mu_R=\xi_R \mz$ and factorisation scale $\mu_F=\xi_F \mz$.
We consider 9 combinations obtained varying $\xi_{R,F} \in [\frac12,1,2]$ and we discard the two cases
$\xi_R=\frac12,\,\xi_F=2$ and $\xi_R=2,\,\xi_F=\frac12$. The percentage correction is computed with respect to the central scales choice
$\xi_R=\xi_F=1$.

Since the parameterisation of the proton must fulfil the sum rules expressing charge and momentum conservation,
then the presence of the photon density in the proton implies a reduction of the total momentum fraction carried by quarks and gluons.
In turn, we observe a reduction of the cross section of the partonic channels induced by quarks and gluons,
which in general should be compensated by the positive cross section of the additional photon-induced channels.
We remark that a similar ${\cal O}(1\%)$ reduction of the quark- and gluon-induced DY cross section for the production of a lepton pair
is balanced by the contribution of the $\gamma\gamma\to \ell^+\ell^-$ process,
of comparable size \cite{CarloniCalame:2007cd,Dittmaier:2009cr,Buonocore:2021bsf}.
In the specific case of on-shell $Z$ production, the $\gamma\gamma$ initial state does not contribute at \oaas, but only at \oaa;
its absence explains, at technical level, the size and sign of the observed effects.

The relevance of the total on-shell $Z$ production cross section as a standard candle for benchmarking purposes is well established.
In the present study, it allows to observe how different PDF collaborations have implemented the QCDxQED evolution
and the photon-density, with a different impact with respect to the corresponding pure QCD analysis.

%% file: conclusion.tex
\section{Conclusions}
We have presented the analytical expression
of the total cross section for the production of an on-shell $Z$ boson at hadron colliders. The result has been expressed in terms of generalised polylogarithmic functions. The three elliptic functions which appear in the double-real contributions have been represented via a series expansion.

These corrections stabilise the prediction of this standard candle, by reducing the size of the uncertainty due to missing higher orders \cite{Bonciani:2020tvf}. The introduction of EW and mixed QCD-EW radiative corrections requires the consistent usage of proton PDFs determined in the same theoretical framework. The comparison with the best prediction obtained in a pure QCD-based model shows that QCD-EW effects up to ${\cal O}(-1\%)$ have to be taken into account and are relevant for the precision determination of this cross section.

\subsection*{Acknowledgments}
We would like to thank F. Caola, M. Delto, P. K. Dhani, M. Jaquier, J.-N. Lang, K. Melnikov, V. Ravindran and R. R\"ontsch for useful discussions. N.R. and A.V. are supported by the Italian Ministero della Universit\`a e della Ricerca (grant PRIN201719AVICI\_01) and by the European Research Council under the European Unions Horizon 2020 research and innovation Programme (grant agreement number 740006). R. B. is partly supported by the italian Ministero della Universit\`a e della Ricerca (MIUR) under grant PRIN 20172LNEEZ. The research of F.B. was partially supported by the ERC Starting Grant 804394 {\sc{hip}QCD}. R.B. and N.R. acknowledge the COST (European Cooperation in Science and Technology) Action CA16201 PARTICLEFACE for partial support.